\documentclass[twocolumn,preprintnumbers,showpacs]{revtex4}
\usepackage{amssymb}
\usepackage{amsmath}
\usepackage{graphicx}

\newcommand{\Lambert}[1]{\mathrm{W}_{#1}}

\newcommand{\dd}{\mathrm{d}}
\newcommand{\ud}{\dd}

\newcommand{\ie}{\textsl{i.e.~}}

\newcommand{\eg}{\textsl{e.g.~}}

\newcommand{\MeV}{\mbox{MeV}}
\newcommand{\GeV}{\mbox{GeV}}
\newcommand{\TeV}{\mbox{TeV}}

\newcommand{\km}{\mbox{km}}
\renewcommand{\sec}{\mbox{s}}

\newcommand{\Mpc}{\mbox{Mpc}}
\newcommand{\bmk}{\boldsymbol k}

\newcommand{\urad}{\mathrm{rad}}
\newcommand{\ureh}{\mathrm{reh}}
\newcommand{\uend}{\mathrm{end}}
\newcommand{\uinf}{\mathrm{inf}}
\newcommand{\unuc}{\mathrm{nuc}}
\newcommand{\uT}{\mathrm{T}}
\newcommand{\uSZ}{\mathrm{SZ}}
\newcommand{\uS}{\mathrm{S}}
\newcommand{\ub}{\mathrm{b}}
\newcommand{\udm}{\mathrm{dm}}

\newcommand{\Mp}{M_{_{\mathrm Pl}}}
\newcommand{\nS}{n_{_{\mathrm S}}}

\newcommand{\Qrms}{Q_{\mathrm{rms-PS}}}

\newcommand{\Rrad}{R_\urad}
\newcommand{\Rreh}{R}
\newcommand{\rhoreh}{\rho_\ureh}
\newcommand{\rhoend}{\rho_\uend}
\newcommand{\rhogamma}{\rho_\gamma}
\newcommand{\rhonuc}{\rho_\unuc}

\newcommand{\Nend}{N_{_\uT}}
\newcommand{\Nreh}{N_\ureh}
\newcommand{\Nstar}{N_*}
\newcommand{\Nzero}{N_0}

\newcommand{\calH}{\mathcal{H}}
\newcommand{\calNnuc}{\mathcal{N}^\unuc}
\newcommand{\calNend}{\mathcal{N}^\uend}
\newcommand{\calFnuc}{\mathcal{F}^\unuc}
\newcommand{\calFend}{\mathcal{F}^\uend}

\newcommand{\OmegaB}{\Omega_\ub}
\newcommand{\OmegaDM}{\Omega_\udm}
\newcommand{\Asz}{A_{_\uSZ}}

\begin{document}

\title{First CMB Constraints on the Inflationary Reheating
  Temperature}

\author{J\'er\^ome Martin} \email{jmartin@iap.fr}
\affiliation{Institut d'Astrophysique de Paris, \\ UMR 7095-CNRS,
Universit\'e Pierre et Marie Curie, \\ 98bis boulevard Arago, 75014
Paris, France}

\author{Christophe Ringeval} \email{christophe.ringeval@uclouvain.be}
\affiliation{Institute of Mathematics and Physics, Centre for Particle
  Physics and Phenomenology, \\ Louvain University, 2 Chemin du
  Cyclotron, 1348 Louvain-la-Neuve (Belgium)}

\date{\today}

\begin{abstract}
  We present the first Bayesian constraints on the single field
  inflationary reheating era obtained from Cosmic Microwave Background
  (CMB) data. After demonstrating that this epoch can be fully
  characterized by the so-called reheating parameter, we show that it
  is constrained by the seven years Wilkinson Microwave Anisotropies
  Probe (WMAP7) data for all large and small field models. An
  interesting feature of our approach is that it yields lower bounds
  on the reheating temperature which can be combined with the upper
  bounds associated with gravitinos production. For large field
  models, we find the energy scale of reheating to be higher than
  those probed at the Large Hadron Collider, $\rhoreh^{1/4}>17.3$ TeV
  at $95\%$ of confidence. For small field models, we obtain the
  two-sigma lower limits $\rho_\ureh^{1/4}>890\,\TeV$ for a mean
  equation of state during reheating $\bar{w}_\ureh=-0.3$ and
  $\rho_\ureh^{1/4}>390\,\GeV$ for $\bar{w}_\ureh=-0.2$.  The physical
  origin of these constraints is pedagogically explained by means of
  the slow-roll approximation. Finally, when marginalizing over all
  possible reheating history, the WMAP7 data push massive inflation
  under pressure ($p<2.2$ at $95\%$ of confidence where $p$ is the
  power index of the large field potentials) while they slightly favor
  super-Planckian field expectation values in the small field models.
 \end{abstract}

\maketitle

\section{Introduction}
\label{sec:intro}

The current ongoing flow of high accuracy astrophysical observations
has important consequences for our understanding of the very early
Universe. In particular, the widely accepted inflationary
paradigm~\cite{Guth:1980zm, Linde:1981mu,
  Albrecht:1982wi,Linde:1983gd} (for a review, see \eg
Refs.~\cite{Linde:2007fr,Martin:2003bt,Martin:2004um,Martin:2007bw})
is now under close scrutiny. According to this scenario, the Cosmic
Microwave Background (CMB) anisotropies and the large scale structures
originate from the unavoidable quantum fluctuations of the inflaton
and gravitational fields in the very early Universe subsequently
amplified during
inflation~\cite{Mukhanov:1981xt,Hawking:1982cz,Starobinsky:1982ee,
  Guth:1982ec,Bardeen:1983qw}. One can show that the corresponding
power spectrum of the cosmological fluctuations naturally acquires an
almost scale invariant form which is fully consistent with all
observations. Another crucial property of the inflationary power
spectrum is that the slight deviations from scale invariance are
linked to the microphysics of
inflation~\cite{Stewart:1993bc,Mukhanov:1990me,Liddle:1994dx}. Therefore,
by measuring these deviations, one can probe the shape of the inflaton
potential and, therefore, learn about the physical origin of the
inflaton field.

\par

It is often claimed from the above properties that observations give
access to a limited part of the potential only, namely the one which
is slow-rolled over by the inflaton when scales of astrophysical
interest today left the Hubble radius. This observational window
represents a range of approximately $7$ e-folds or three decades in
wavenumbers. However, inflation does not consist of the slow-roll
phase only and the pre and/or reheating period is also of fundamental
importance since it allows us to understand how inflation is connected
to the hot big-bang
phase~\cite{Turner:1983he,Kofman:1997yn,Bassett:2005xm,
  Mazumdar:2010sa}. This physical phenomenon is related to a different
part of the inflationary potential, usually the one located close to
its true minimum, \ie a few decades in e-folds away from the
observable window.

\par

Observation of pre-/reheating effects can be achieved in two
ways. First, the power spectrum can evolve on large scales when the
inflaton field oscillates around the minimum of its
potential. However, this happens only in quite complicated models,
typically those containing more than one field~\cite{Finelli:1998bu,
  Bassett:1998wg, Finelli:2000ya}. In fact, it was recently shown that
this type of effect can also happen in single field inflation but on
much smaller
scales~\cite{Jedamzik:2010dq,Jedamzik:2010hq,Easther:2010mr}. Second,
the duration of the pre-/reheating phase can significantly modify the
position of the observational window mentioned above. Put differently,
at fixed astrophysical scales today, changing the pre-/reheating
duration is equivalent to moving the window along the potential, hence
probing different values of the power spectrum spectral index,
amplitude of the anisotropies and tensor-to-scalar ratio. Obviously,
this cannot be done arbitrarily because CMB data impose accurate
bounds on their value. Conversely, this opens up the possibility to
constrain the pre-/reheating duration and/or its equation of state
from CMB data~\cite{Martin:2006rs}. Notice that a direct detection of
primordial gravitational waves would also allow us to probe the
reheating temperature, as shown in Refs.~\cite{Nakayama:2008wy,
  Kuroyanagi:2009br}.

\par

The goal of this article is to address this question for the standard
scenarios of inflation. It is traditional to study three categories of
models usually considered as representative of the full inflationary
space. These models are large field~\cite{Linde:1984st}, small
field~\cite{Linde:1981mu,Albrecht:1982wi}, and hybrid
inflation~\cite{Linde:1993cn}. Hybrid scenarios involve multiple
fields and, therefore, the power spectrum can change during the
preheating phase. This makes this class of scenarios deserving of a
separate investigation. For this reason, in this article, we limit
ourselves to the class of large and small field models.

\par

In the following, we will use the term ``reheating'' to refer to the
pre-/reheating phases of the Universe defined to have occurred just
after the end of inflation and just before the radiation dominated
era. So far, the constraints on the reheating energy scale are not so
numerous. Obviously, it should be less than the energy scale of
inflation which implies that $T_\ureh\lesssim 10^{16}$ GeV. In
addition, if one assumes that supersymmetry is the correct extension
of the standard model of particle physics, then constraints from
Big-Bang Nucleosynthesis (BBN) on unstable gravitinos lead to a
reheating temperature $T_\ureh\lesssim 10^7$ GeV~\cite{Khlopov:1984pf,
  Kallosh:1999jj, Giudice:1999yt, Lemoine:1999sc, Maroto:1999ch,
  Giudice:1999am, Buonanno:2000cp, Copeland:2005qe,Jedamzik:2006xz,
  Kawasaki:2008qe, Bailly:2009pe}. Notice that this constraint can
nevertheless be avoided if one considers the scenario of
Ref.~\cite{Mardon:2009gw}. Reheating itself should also proceed before
BBN and this implies that $T_\ureh \gtrsim 10$ MeV. We see that the
reheating temperature is poorly constrained, in particular, its lower
limit. As a matter of fact, the work presented here precisely yields a
lower limit on the reheating energy scale from the current seven years
Wilkinson Microwave Anisotropies Probe (WMAP7)
data~\cite{Jarosik:2010iu, Komatsu:2010fb, Larson:2010gs}.

\par 

In order to derive constraints on the reheating phase, we make use of
Bayesian techniques and utilize a full numerical
approach~\cite{Ringeval:2007am}. This has several advantages. First,
it is exact and rests only on the linear theory of cosmological
perturbations: the method remains accurate when the slow-roll
approximation breaks down, as one expects near the end of
inflation. Second, and of particular importance for the present work,
it permits a new treatment of reheating. Indeed, instead of viewing
the reheating parameters as nuisance parameters, they can easily be
included in the Bayesian data analysis process. Third, the evolution
of cosmological perturbations in the hot big-bang eras already relies
on numerical codes. Treating perturbations during inflation in the
same way allows the whole procedure to be automatized and to be easily
extended to other scenarios. Fourth, the numerical approach allows us
to address the question of the priors choice in a particularly
well-defined way. Indeed, from a physical point of view, our prior
knowledge is on the inflationary theory and not on the shape of the
primordial power spectra which is actually a model
prediction. Therefore, it is better, and easier, to choose prior
probability distributions directly on the model parameters, such as
the power index of the large field potentials. This reflects the fact
that a model of inflation is not a disembodied mathematical structure
that one only needs to ``fit'' but a physical scenario rooted in high
energy physics that one needs to understand.

\par

This paper is organized as follows. In Sec.~\ref{sec:physicalorigin},
we extend the above discussion and explain in detail why the reheating
epoch can be constrained with CMB data. In particular, we introduce
the so-called reheating parameter which depends on the reheating
duration and on the mean equation of state of the fluid dominating the
Universe during this epoch. Then, using the slow-roll approximation,
we analytically demonstrate that the accuracy of the WMAP7 data is now
sufficient to obtain some constraints on the reheating era. In
Sec.~\ref{sec:cmbtoreh}, using a full numerical integration of the
tensor and scalar power spectra coupled to Bayesian methods, we derive
the constraints that any reheating model has to satisfy. Then,
assuming specific values for the mean equation of state, we translate
these constraints into new lower limits for the reheating energy
density and/or reheating temperature. These results significantly
improve the bounds coming from the Big-Bang Nucleosynthesis. In
Sec.~\ref{sec:conclusion}, we recap our main findings and discuss how
our results are modified by the inclusion of others CMB data sets. In
Appendix~\ref{appendix:lfreh}, we work out a typical example which
illustrates the robustness of our assumptions: a noninstantaneous
transition between reheating and the radiation dominated era when one
considers the finite decay width of the inflaton field. Finally, as a
by-product of our data analysis, Appendix~\ref{appendix:srpost}
presents the updated WMAP7 constraints on the spectral index,
tensor-to-scalar ratio and first order slow-roll parameters
marginalized over second order effects.

\section{Physical origin of the constraint}
\label{sec:physicalorigin}

Before presenting and discussing the constraints on the reheating
temperature, we explain why and how these ones can be inferred from
high accuracy CMB observations. In particular, we use the slow-roll
approximation to explicitly illustrate the method.

\subsection{Parametrizing the reheating}
\label{subsec:parameter}

The evolution of scalar (density) perturbations is controlled by the
so-called Mukhanov--Sasaki variable $v_{{\bmk}}$. If matter is
described by a scalar field (as is the case during inflation and
pre-/reheating), then its equation of motion is given, in Fourier
space,
by~\cite{Mukhanov:1990me,Martin:2003bt,Martin:2004um,Martin:2007bw}
\begin{equation}
\label{eq:eqmotv}
v_{{\bmk}}''+\left[k^2-\frac{\left(a\sqrt{\epsilon_1}\right)''}
{a\sqrt{\epsilon_1}}\right]v_{{\bmk}}=0.
\end{equation}
Here, a prime denotes a derivative with respect to conformal time. The
quantity $k$ is the comoving wave number and $\epsilon _1\equiv
-\dot{H}/H^2$ is the first Hubble flow function~\cite{Schwarz:2001vv},
$H=\dot{a}/a$ being the Hubble parameter and $a$ the
Friedmann--Lema\^{\i}tre--Robertson--Walker (FLRW) scale factor (a dot
means derivative with respect to cosmic time). The quantity $v_{\bmk}$
is related to the curvature perturbation $\zeta_{\bmk}$ through the
following expression:
\begin{equation}
\label{eq:zetavsv}
\zeta_{\bmk}=\frac{1}{\Mp}\frac{v_{\bmk}}{a\sqrt{2\epsilon
    _1}}\,,
\end{equation}
where $\Mp$ stands for the reduced Planck mass. As a consequence, the
power spectrum of $\zeta _{\bmk}$ can be expressed as
\begin{equation}
{\cal P}_{\zeta}(k)\equiv \frac{k^3}{2\pi ^2}
\left\vert \zeta_{\bmk}\right\vert ^2
=\frac{k^3}{4\pi^2 \Mp^2}\left\vert 
\frac{v_{\bmk}}{a\sqrt{\epsilon _1}}\right\vert ^2\, .
\label{Pzeta}
\end{equation}
In order to calculate ${\cal P}_{\zeta}(k)$, one needs to integrate
Eq.~(\ref{eq:eqmotv}), which requires the knowledge of the initial
conditions for the mode function $v_{\bmk}$. Since, at the beginning
of inflation, all the modes of astrophysical interest today were much
smaller than the Hubble radius, the initial conditions are chosen to
be the Bunch-Davis vacuum which amounts to
\begin{equation}
\lim_{k/\calH \rightarrow +\infty}v_{\bmk}=\frac{1}{\sqrt{2k}}
{\rm e}^{-ik\eta }\, ,
\label{eq:initial}
\end{equation}
where $\eta $ denotes conformal time and $\calH=aH$ is the conformal
Hubble parameter. The importance of the curvature perturbation lies in
the fact that it is directly related to CMB anisotropies, the two
point correlation function of which can be expressed in term of the
spectrum of $\zeta _{\bmk}$. Moreover, under very general conditions
(including the assumption that inflation proceeds with only one
field), $\zeta _{\bmk}$ is a conserved quantity on large scales and,
therefore, can be used to propagate the inflationary spectrum from the
end of inflation to the post-inflationary era~\cite{Martin:1997zd}. In
other words, the power spectrum is not affected by the
post-inflationary evolution, in particular by the pre-/reheating
epoch.

\par

However, this does not mean that the reheating era has no effect on
the inflationary predictions. On the contrary, the relation between
the physical scales at present time and during inflation depends on
the properties of this phase of evolution. As a consequence, in order
to calibrate the inflationary spectrum with respect to the physical
scales of astrophysical interest today, it is necessary to know how
the reheating phase proceeded. Conversely, this also opens the
possibility to constrain the physical conditions that prevailed at
that time by means of CMB observations.

\par

In order to put the above considerations on a quantitative footing,
let us rewrite Eq.~(\ref{eq:eqmotv}) in terms of the number of e-folds
during inflation, $N\equiv \ln\left(a/a_{\rm ini}\right)$, where
$a_{\rm ini}$ is the value of the scale factor at the beginning of
inflation. It takes the form
\begin{equation}
\label{eq:eomNefold}
  \frac{{\rm d}^2v_{\bmk}}{{\rm d}N^2}+\frac{1}{\cal H}
  \frac{{\rm d}{\cal H}}{{\rm d}N}\frac{{\rm d}v_{\bmk}}{{\rm d}N}
  +\left[\left(\frac{k}{\cal H}\right)^2
    -U_{_\uS}(N)\right]v_{\bmk}=0,
\end{equation}
where $U_{_\uS}(N)$ is an effective potential for the perturbations
which depends on the scale factor and its derivatives only. All the
terms in this equation but $k/{\cal H}$ are completely specified by
the inflationary background evolution. In practice, we are given a
physical scale today, say $k/a_{\rm now}$ (for instance $k/a_{\rm now}
= 0.05\, \Mpc^{-1}$) and we need to express $k/{\cal H}$ in terms of
$k/a_{\rm now}$ and quantities defined during
inflation. Straightforward considerations lead to
\begin{equation}
\frac{k}{{\cal H}}=\frac{\Upsilon_{\bmk}}{H(N)}{\rm e}^{N_{_{\rm T}}-N},
\end{equation}
where $N_{_{\rm T}}$ is the total number of e-folds during inflation
and $\Upsilon _{\bmk}$ is defined by
\begin{equation}
  \Upsilon_{\bmk}\equiv \frac{k}{a_{\rm now}}\left(1 + z_\uend\right),
\end{equation}
with $z_\uend$ being the redshift of the end of inflation. As expected
$\Upsilon _{\bmk}$ depends on the whole post-inflationary history
through $z_\uend$. During this post-inflationary history, only the
reheating phase is poorly known and represents, by far, the main
source of uncertainty for the inflationary predictions. For
convenience, we rewrite $\Upsilon_{\bmk}$ as
\begin{equation}
\label{eq:defRrad}
  \Upsilon_{\bmk}=\frac{k}{a_{\rm now}}
  \left(\frac{\rho_{\rm end}}{\Omega _{\gamma}\rho_{\rm cri}}\right)^{1/4}
R_{\rm rad}^{-1},
\end{equation}
thus defining the new parameter $R_{\rm rad}$. This parameters plays a
crucial role in this article. In the above equation, $\rho_{\rm end}$
is the energy density at the end of inflation, $\rho_{\rm cri}$ is the
present day critical energy density and $\Omega _{\gamma}\simeq
2.471\times 10^{-5}h^{-2}$ is the density parameter of radiation
today. As a result $\Omega _{\gamma }\rho_{\rm cri}\equiv \rho_\gamma$
is the present day radiation energy density and does not depend on
$h^2$. The above equations make clear that the parameter $\Rrad$ must
be specified if one wants to compare an inflationary model to
observations.

\par

In fact, the quantity $\Rrad$ has a simple physical
interpretation. Let us assume that the reheating phase is dominated by
a conserved effective fluid with energy density $\rho$ and pressure
$P$. The fact that we assume the effective fluid to be conserved is
not a limitation. For instance, in a simple model where the inflaton
scalar field is coupled to radiation (see
Appendix~\ref{appendix:lfreh}), the effective fluid is just defined by
$\rho=\rho _{\phi}+\rho _{\gamma }$ and $P = P_{\phi}+\rho_{\gamma
}/3$. The scalar field and the radiation are not separately conserved
but the effective fluid is. Then, it is straightforward to show that
\begin{equation}
\label{eq:rhorehend}
\rho(N)=\rhoend
\exp\left\{-3\int _{\Nend}^{N}\left[1+w_{\ureh}(n)\right]{\ud} n\right\},
\end{equation}
where $w_{\ureh}\equiv P/\rho$ is the equation of state function
during reheating. Using this expression, one obtains
\begin{equation}
\label{eq:Rrad}
\ln \Rrad = \frac{\Delta N}{4}\left(-1+3\bar{w}_{\ureh}\right),
\end{equation}
where
\begin{equation}
  \Delta N\equiv N_{\ureh} - \Nend,
\end{equation}
is the total number of e-folds during reheating, $N_{\ureh}$ being the
number of e-folds at which reheating is completed and the radiation
dominated era begins. The quantity $\bar{w}_{\ureh}$ stands for the
mean equation of state parameter
\begin{equation}
\bar{w}_{\ureh}\equiv \frac{1}{\Delta N}\int _{N_{_{\rm T}}}^{N_{\ureh}}
w_{\ureh}(n){\rm d}n.
\end{equation}
Therefore, the parameter $R_{\rm rad}$ only depends on what happens
during reheating. To put differently, it singles out in the
expression of $\Upsilon_{\bmk}$, the contribution coming from
reheating. Equation~(\ref{eq:Rrad}) also allows us to understand
why $R_{\rm rad}$ carries the subscript ``rad''. Indeed, if the
effective fluid is equivalent to radiation, then $\bar{w}_\ureh=1/3$
and $\ln \Rrad=0$. The physical interpretation is very clear: in this
case the reheating stage cannot be distinguished from the subsequent
radiation dominated era and, therefore, cannot affect the inflationary
predictions: as a consequence $\Rrad=1$ in Eq.~(\ref{eq:defRrad}).

\par

In fact, one can even go further and express $\Rrad$ in an even more
compact form. Using Eq.~(\ref{eq:rhorehend}), one can write
$\rhoreh=\rhoend \exp\left[-3\Delta N(1+\bar{w}_{\ureh})\right]$ from
which, together with Eq.~(\ref{eq:Rrad}), one obtains
\begin{equation}
\label{eq:Rradw}
\ln R_{\rm rad}=\frac{1-3\bar{w}_{\ureh}}{12(1+\bar{w}_{\ureh})}\ln \left(
\frac{\rhoreh}{\rhoend}\right),
\end{equation} 
where $\rhoreh$ has to be understood as the energy density at the end
of the reheating era, \ie $\rho(\Nreh)$.

\par

Let us summarize our discussion. In order to calculate the power
spectrum of the inflationary cosmological perturbations, one needs to
solve Eq.~(\ref{eq:eomNefold}). In this formula, all the terms are
accurately known during inflation except
\begin{equation}
  \frac{k}{\cal H}=\frac{k}{a_{\rm now}}
\left(\frac{\rhoend}{\rho_\gamma}\right)^{1/4}\frac{1}{H(N)\Rrad}
{\rm e}^{N_{_{\rm T}}-N},
\end{equation}
and the theoretical uncertainty in this expression solely comes from
the parameter $\Rrad$ which depends on reheating only (more precisely,
on the energy density at the end of reheating, $\rhoreh$, and the mean
equation of state $\bar{w}_{\ureh}$).

\subsection{Why CMB observations constrain reheating}
\label{subsec:why}

Having discussed the physical interpretation of $R_{\rm rad}$, we now
explain how the CMB observations can constrain its value. For this
purpose, we reexpress $R_{\rm rad}$ in terms of quantities defined at the
Hubble radius crossing. One obtains
\begin{eqnarray}
\label{eq:Rradsr}
\ln R_{\rm rad} &=& N_{_{\rm T}}-N_*+N_0
-\frac14\ln\left(\frac{H_*^2}{\Mp^2\epsilon_{1*}}\right)
\nonumber \\ &+& 
\frac{1}{4}\ln\left(\frac{3}{\epsilon_{1*}}
\frac{V_{\rm end}}{V_*}\frac{3-\epsilon_{1*}}
{3-\epsilon_{1\,\rm end}}\right),
\end{eqnarray}
where we have defined
\begin{equation}
  N_0 \equiv \ln \left( \dfrac{k/a_{\rm now}}{\rhogamma^{1/4}} \right).
\end{equation}
In this formula, $N_*$ is the e-folds number at which the scale
$k/a_{\rm now}$ crossed out the Hubble radius during inflation (all
the quantities with a subscript ``*'' are evaluated at that time) and
$V(\phi)$ is the inflaton potential. Despite the appearance of the
first Hubble flow function, this equation is exact (moreover, we also
have $\epsilon_{1 \rm end}=1$). At leading order, one has
\begin{equation}
\dfrac{H_*^2}{\Mp^2\epsilon_{1*}} = 8\pi^2 P_*,
\end{equation}
where the amplitude of the scalar power spectrum at the pivot scale
$P_* = {\cal P}_{\zeta}(k_*)$ is directly related to the Cosmic
Background Observer (COBE) normalization.

\par

The above equation~(\ref{eq:Rradsr}) can be used in two different
manners. The first way is to assume something about $\Rrad$ and to
derive the corresponding range of variations of the inflationary
slow-roll predictions $N_*$ and $\epsilon_i(N_*)$. In other words,
this determines how the inflationary predictions depend on the details
of the reheating era. This approach is the one usually considered in
the literature to compare inflationary predictions to the current
constraints on the slow-roll parameters $\epsilon_{i*}$ (or spectral
index and tensor-to-scalar ratio). Unfortunately, the assumptions on
$\Rrad$ are rarely explicit and comparison is only made by choosing
reasonably assumed values of $N_*$: typically $30$ and $60$ e-folds as
one may derive under generic
assumptions~\cite{Liddle:2003as}. However, as Eq.~(\ref{eq:Rradsr})
explicitly shows, once $V(\phi)$ is chosen, and the tilt and amplitude
of the scalar perturbations measured, $N_*$ is directly related to
$\Rrad$, which itself, as already noticed, depends on the energy
density $\rhoreh$ at which reheating ends and $\bar{w}_{\ureh}$. As a
result, the range of variation for $N_*$ can only be known once a
reheating model is assumed. Without such an assumption, from one model
to another, an assumed value of $N_*$ may inconsistently imply that
the reheating occurs after nucleosynthesis, or even at energy
densities higher than $\rhoend$. This type of model would therefore
appears to be compatible with the CMB data favored power spectra while
being totally inconsistent with standard cosmology.

\par

Let us now see how it works in practice. In order to be consistent
with the standard cosmological model, $\ln \Rrad $ cannot take
arbitrary values. One should have $\bar{w}_{\ureh}<1$ to respect the
positivity energy conditions of General Relativity and $\bar{w}_\ureh
> -1/3$ by the very definition of reheating which is not
inflation. Notice that we impose conditions on the mean value of the
equation of state only. In addition, reheating should occur after
inflation and before BBN, \ie $\rho _{\rm nuc}<\rho _{\ureh}
<\rho_{\rm end}$, with
\begin{equation}
\rhonuc \equiv \left(10 \MeV \right)^4.
\end{equation}
This allows us to explicitly use Eq.~(\ref{eq:Rradsr}). Combined with
Eq.(\ref{eq:Rradw}), we can determine the range of variation of
$\Delta N_*\equiv N_{_{\rm T}}-N_*\in [\Delta N_*^{\rm nuc},\Delta
N_*^{\rm end}]$. Straightforward manipulations lead to
\begin{eqnarray}
\label{eq:Nnuc}
\Delta N_*^{\rm nuc}&=&-N_0+
\ln \left(\frac{H_*}{\Mp}\right)- 
\frac{1}{3(1+\bar{w}_{\ureh})}\ln \frac{\rho_{\rm end}}{\Mp^4}
\nonumber \\ &&
+\frac{1-3 \bar{w}_\ureh}{12(1+\bar{w}_{\ureh})}
\ln \frac{\rho_{\rm nuc}}{\Mp^4}\, ,
\end{eqnarray}
while if one chooses $\rho_{\ureh}=\rho_{\rm end}$, one obtains
\begin{eqnarray}
\label{eq:Nend}
\Delta N_*^{\rm end}&=&-N_0+
\ln \left(\frac{H_*}{\Mp}\right)
-\frac{1}{4}\ln \frac{\rho_{\rm end}}{\Mp^4}.\nonumber \\
\end{eqnarray}
Interestingly enough, the last equation no longer depends on
$\bar{w}_{\ureh}$. This is of course because requiring
$\rhoreh=\rhoend$ means that one immediately reheats the Universe
after inflation. It is important to notice that these equations are
algebraic for $\Delta N_*^{\unuc}$ and $\Delta N_*^{\uend}$ because
$H_*$ and $\rhoend$ are also functions of $\Delta N_*$. The
corresponding range of variations of the inflationary predictions is
determined by calculating $\Delta \epsilon_{i*}=\epsilon_i(\Delta
N_*)$ with $\Delta N_*$ given above. To proceed further, one needs to
specify the model of inflation. In the next section, one considers the
prototypical scenario of chaotic inflation as well as the small field
models.

\par

Before dealing with these explicit examples, let us briefly anticipate
and discuss the second way of using Eq.~(\ref{eq:Rradsr}). It consists
of considering $\Rrad$ as an observable model parameter and in
including it in the data analysis, as it should be from a Bayesian
point of view. If we are given a specific potential, then $V_\uend$ is
explicitly known. CMB data put a limit on $H_*^2/\epsilon_{1*}$
through the amplitude of the anisotropies, as well as on
$\epsilon_{1*}$ from the tensor-to-scalar ratio. As a result, one
expects CMB data to also give some information on $\Rrad$. This is the
subject of Sec.~\ref{sec:cmbtoreh} in which we perform a Bayesian data
analysis of the WMAP7 data for both the large and small field models
by including the reheating. For the first time, we find that $\Rrad$
is not only a nuisance parameter for inflation but ends up being
constrained by the WMAP7 data. We then discuss the physical
implications of these bounds and show that CMB data give us a lower
bound on the energy scale at which the reheating ended.

\subsection{Large field models}
\label{subsec:largefield}

We now consider the archetypal model of inflation, namely, large
field inflation. This working example is important because it allows
us to show explicitly which type of constraints one should
expect. Large field models are characterized by the potential
\begin{equation}
\label{eq:lfpot}
V(\phi)=M^4\left(\frac{\phi}{\Mp}\right)^p,
\end{equation}
where $M$ is an energy scale which fixes the amplitude of the CMB
anisotropies and $p$ is a free index. In this case the slow-roll
trajectory is explicitly known and one can calculate $\phi_*$, the
field vacuum expectation value at Hubble radius crossing from
$\phi_{\uend}/\Mp=p/\sqrt{2}$, the field vacuum expectation value
(VEV) at which inflation stops. One gets~\cite{Martin:2006rs}
\begin{equation}
\phi_*^2 = 2 p \Mp^2\Delta N_* + \phi_\uend^2.
\end{equation}

\par

The reheating phase in large field models proceeds by parametric
oscillations around the minimum of the potential and it is well known
that the corresponding equation of state parameter is given
by~\cite{Turner:1983he, Kofman:1997yn, Liddle:2003as, Martin:2003bt}
\begin{equation}
\label{eq:wrehlf}
  \bar{w}_{\ureh}=\dfrac{p-2}{p+2}\,.
\end{equation}
In particular, for $p=2$, one obtains $\bar{w}_{\ureh}=0$: that is to
say the oscillatory phase is equivalent to a matter dominated era (the
quartic case corresponding to a radiation dominated era, and so
on). Although this formula is derived without taking into account the
coupling between the inflaton field and radiation, we show in
Appendix~\ref{appendix:lfreh} that it is a very good approximation.

\par

Knowing explicitly the equation during reheating, we are now in a
position where the algebraic equations~(\ref{eq:Nnuc})
and~(\ref{eq:Nend}) can be solved exactly. After some algebra, one
obtains
\begin{eqnarray}
\label{eq:Dnnuclf}
\Delta N_*^{\rm nuc}&=&-\frac{p}{4}-\frac{p^2-2p+4}{12p}\Lambert{0}
\biggl\{-\frac{12p}{p^2-2p+4}
\nonumber \\ & & \times
\exp\left[-\frac{12p\left({\cal N}^{\rm nuc}+p/4\right)}
{p^2-2p+4}\right]\biggr\}, 
\end{eqnarray}
and
\begin{eqnarray}
\label{eq:Dnendlf}
\Delta N_*^{\rm end}&=&-\frac{p}{4}-\frac{p-2}{8}\Lambert{0}
\Biggl\{-\frac{8}{p-2}
\nonumber \\ & & \times
\exp\left[-\frac{8\left({\cal N}^{\rm end}+p/4\right)}
{p-2}\right]\Biggr\},
\end{eqnarray}
where $\Lambert{0}$ is a Lambert function. Both quantities $\calNnuc$
and $\calNend$ depend only on the model parameter $p$ and the
amplitude of the observed anisotropies.  Explicitly, $\calNnuc$ reads
\begin{equation}
\label{eq:calnuc}
\begin{aligned}
  \calNnuc &= -N_0
  +\dfrac{2}{3p}(p-1)\ln\left(2\pi\sqrt{120}\frac{\Qrms}{T}\right) \\
    & -\frac{p+2}{6p}\ln\left[9 \frac{2^{(-p^2+p-6)/(p+2)}}
      {p^{(-p^2+2p-4)/(2p+4)}}\right] -\frac{p-4}{3p}\ln
    \frac{\rhonuc}{\Mp}\,,
\end{aligned}
\end{equation}
where the amplitude of the CMB anisotropies has been expressed in
terms of the quadrupole moment
\begin{equation}
\dfrac{\Qrms}{T} = \sqrt{\dfrac{5 C_2}{4\pi}} \simeq 6\times 10^{-6}.
\end{equation}
In Eq.~(\ref{eq:calnuc}), the last term vanishes for $p=4$ since, as
already noticed above, the phase of oscillations is equivalent to a
radiation dominated era which cannot be distinguished from the
subsequent hot big-bang epoch. On the other hand, the constant
$\calNend$ can be expressed as
\begin{equation}
\label{eq:calend}
\begin{aligned}
  \calNend =
  -N_0 &+\frac12\ln\left(2\pi\sqrt{120}\frac{\Qrms}{T}\right) \\ 
  &- \frac{1}{4}\ln \left(9\frac{2^{1-p}}{p^{1-p/2}}\right).
\end{aligned}
\end{equation}

\begin{figure*}
\begin{center}
\includegraphics[width=0.95\textwidth,clip=true]{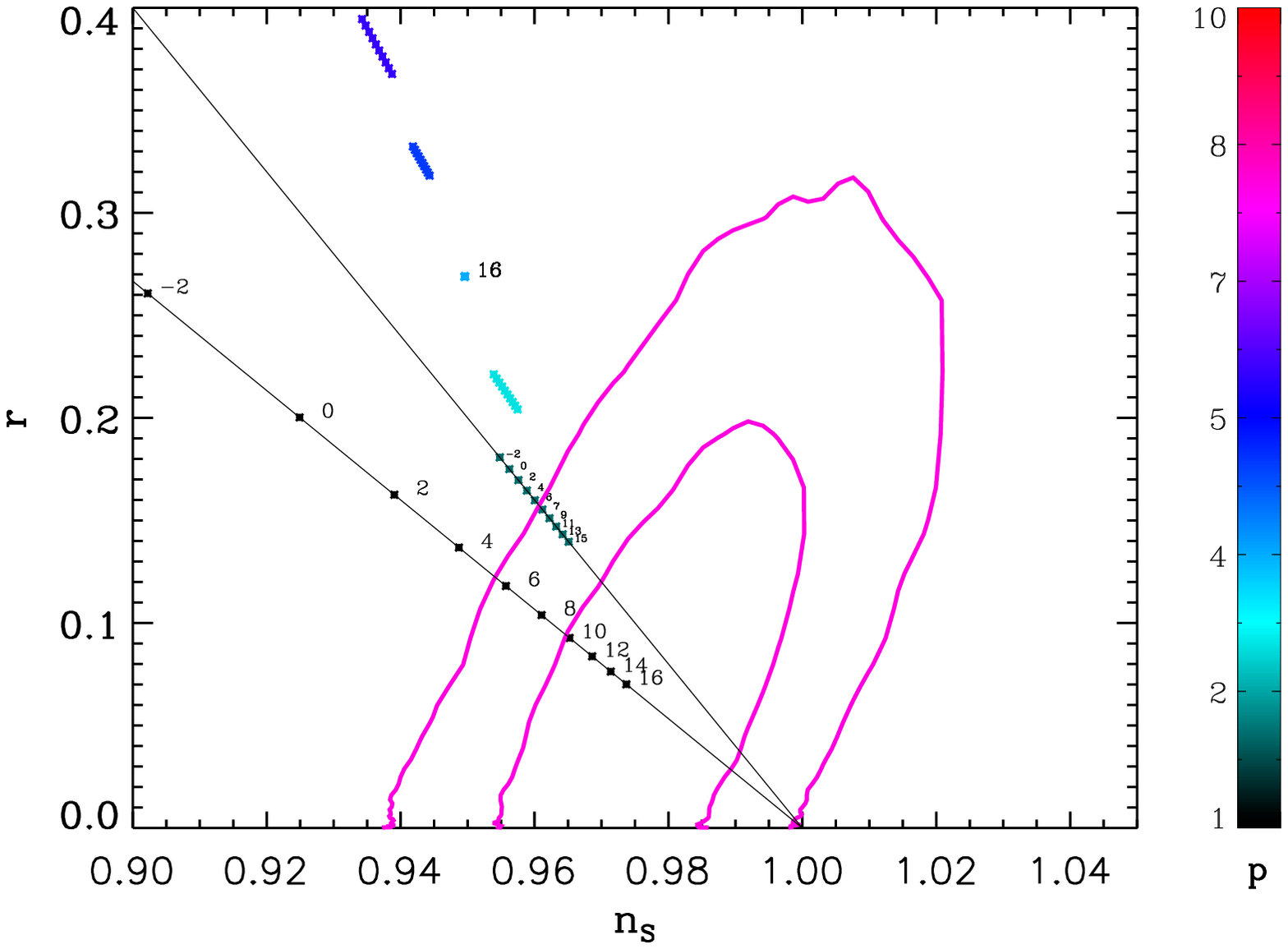}
\caption{Reheating consistent slow-roll predictions for the large
  field models in the plane $(\nS,r)$. The two contours are the one
  and two-sigma WMAP confidence intervals (marginalized over second
  order slow-roll). The two lines represent the locus of the $p
  \gtrsim 1$ and $p=2$ models while the blue point annotated ``$16$''
  corresponds to $p=4$. The annotations trace the energy scale at
  which the large field reheating ends and correspond to
  $\log(g_*^{1/4}T_{\ureh}/\GeV)$. Clearly, these values are limited
  from below to stay inside the two-sigma contours.}
\label{fig:srlfnsR}
\end{center}
\end{figure*}

\begin{figure*}
\begin{center}
\includegraphics[width=0.95\textwidth,clip=true]{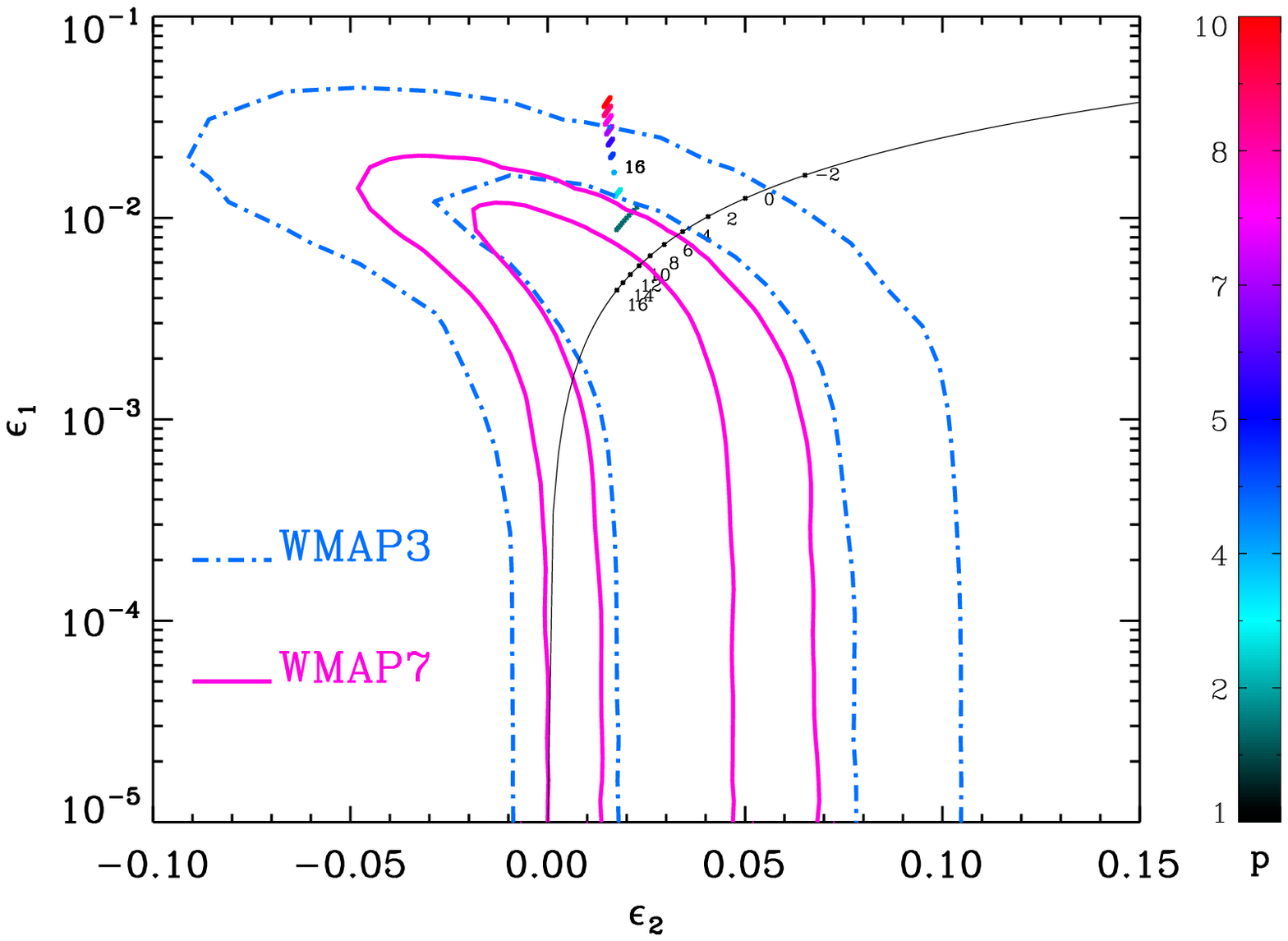}
\caption{Reheating consistent slow-roll predictions for the large
  field models in the plane $(\epsilon_1,\epsilon_2)$. The two blue
  dot-dashed contours are the one- and two-sigma WMAP3
  confidence intervals (marginalized over second order slow-roll)
  while the pink solid contours are the one- and two-sigma WMAP7
  ones. As in Fig.~\ref{fig:srlfnsR}, the annotations trace the
  energy scale at which the large field reheating ends and correspond
  to $\log(g_*^{1/4}T_{\ureh}/\GeV)$. The solid line represents the
  model $p \gtrsim 1$. This confirms that there now exists a lower
  bound on the value of $g_*^{1/4}T_{\ureh}$ (see
  Sec.~\ref{sec:cmbtoreh}).}
\label{fig:srlf}
\end{center}
\end{figure*}

{}From Eqs.~(\ref{eq:Dnnuclf}) and (\ref{eq:Dnendlf}), we immediately
deduce that, for the large field models, the range of allowed values
of $\Delta N_*$ strongly depends on $p$. In Figs.~\ref{fig:srlfnsR}
and~\ref{fig:srlf}, we have plotted the large field predictions,
obtained from Eqs.~(\ref{eq:Dnnuclf}) and (\ref{eq:Dnendlf}), for the
slow-roll parameters $\epsilon_{1}(\Delta N_*)$ and
$\epsilon_{2}(\Delta N_*)$ compared to the one- and two-sigma WMAP7
confidence intervals (see Appendix~\ref{appendix:srpost}). The first
figure represents the slow-roll predictions in the plane $(\nS,r)$,
while the second one corresponds to the plane
$(\epsilon_1,\epsilon_2)$. The annotated values trace the quantity
$\log(g_*^{1/4}T_{\ureh}/\GeV)$, where the reheating temperature is
defined by the relation
\begin{equation}
\label{eq:Trehdef}
g_*^{1/4}T_{\ureh}\equiv \left(\dfrac{30}{\pi^2} \rhoreh \right)^{1/4},
\end{equation}
and $g_*$ is the number of relativistic degrees of freedom at that
time. Large values of $p$ cannot explain the current measurements of
$\nS$ and $r$, while the low values can. Therefore, it is clear that
there now exists a lower bound on the reheating temperature. In
Sec.~\ref{sec:cmbtoreh}, we derive the Bayesian two-sigma limits on
$\rhoreh$ (or $g_*^{1/4}T_{\ureh}$) by including the reheating
parameter into the data analysis process. These plots should make
evident that the reheating in the large field models is already
observable with the current CMB data, and more than being a nuisance
parameter, it is actually constrained.

\par

Let us also remark that the case $p=4$ is particularly interesting;
see the blue point annotated ``$16$'' in Figs.~\ref{fig:srlfnsR}
and~\ref{fig:srlf}. Indeed, the value $p=4$ is the extreme case in
which $\Delta N_*$ is actually fixed to
\begin{equation}
  \Delta N_*^{p=4} = -\Nzero + \dfrac{1}{2}
  \ln \left( 2\pi\sqrt{120} \frac{\Qrms}{T}\right) \simeq 58.5\,.
\end{equation}
This is why this model is represented by a single point in
Figs.~\ref{fig:srlfnsR} and~\ref{fig:srlf}. Making any other choice is
equivalent to assuming a more complicated reheating model which should
at least be specified. For instance, Ref.~\cite{Komatsu:2010fb} (see
Fig.~19) uses two values, $50$ and $60$ e-folds, instead of one. From
the above considerations, it is clear that $50$ is much too
small. But, of course, one can always assume that the shape of the
potential in the slow-roll regime is not the same as in the reheating
regime (actually, this has to be the case for small field models, see
below). In this case $V(\phi)\propto \phi^4$ is not relevant during
the oscillations of the field and $\bar{w}_{\ureh}\neq 1/3$. However,
as discussed in the next section, even in this case, the reheating
epoch is still constrained. The same range of variations for $\Delta
N_*$ has also been used in Ref.~\cite{Finelli:2009bs} (see Fig.~2) for
the $\phi^2$ model. Compared to our results, $50$ is too high and
excludes models which are still allowed while $60$ predicts a reheating
energy scale higher the energy scale at the end of inflation:
$\rho_\ureh> \rho_\uend$.

\subsection{Small field models}
\label{subsec:smallfield}

One of the reasons leading to such a strong reheating influence on the
large field model predictions comes from Eq.~(\ref{eq:wrehlf}). Once
the potential is chosen, the spectral index and tensor-to-scalar ratio
are intimately linked to the way the reheating proceeds. One may
therefore wonder how the reheating can influence the model predictions
in a case where it is unrelated to the shape of the primordial power
spectra. As a motivated example, we discuss in this section the case
of the small field models ending with a reheating characterized by a
mean equation of state $\bar{w}_\ureh$. The small field potential
reads
\begin{equation}
\label{eq:sfpot}
V(\phi) = M^4 \left[1- \left( \dfrac{\phi}{\mu} \right)^p \right],
\end{equation}
where $\mu$ represents a VEV for the field $\phi$, $p$ a power
index, and $M$ fixes the amplitude of the observed
anisotropies. Inflation proceeds from small to large values of the
field. For convenience, we denote by $\chi$ the field value in units of
$\mu$, \ie $\chi \equiv \phi/\mu $. The slow-roll trajectory in terms
of $\chi$ reads~\cite{Martin:2006rs}
\begin{equation}
\label{eq:sftraj}
  \Delta N_* = \dfrac{\mu^2}{2p \Mp^2} \left(\chi_*^2 + 
\dfrac{2}{p-2} \chi_*^{2-p} - \chi_\uend^2 + \dfrac{2}{2-p} 
\chi_\uend^{2-p} \right),
\end{equation}
where, again, $\chi_\uend$ is defined by $\epsilon_1(\chi_\uend) = 1$, \ie
\begin{equation}
\label{eq:chiend}
  \chi_\uend^{p-1} = \dfrac{\sqrt{2}}{p} \dfrac{\mu}{\Mp}
  \left(1 - \chi_\uend^p \right).
\end{equation}
Both of these equations do not have an explicit solution (unless
$p=2$) but can be numerically solved for a given set of model
parameters $\mu$ and $p$. For this reason, instead of deriving the
reheating allowed values for $\Delta N_*$, it is more convenient to
derive the reheating allowed values for $\chi_*$. In fact, $\chi_*$
should lie between $\chi_*^\unuc$ and $\chi_*^\uend$, the field values
such that reheating ends, respectively, at BBN and just after
inflation. After some algebra, one finds $\chi_*^\unuc$ to be the
solution of
\begin{equation}
\label{eq:chistarnuc}
\begin{aligned}
  \dfrac{\mu^2}{2p \Mp^2} & \left(\chi_* + \dfrac{2}{p-2} \chi_*^{2-p}
  \right) + \dfrac{3 \bar{w}_\ureh}{3 + 3 \bar{w}_\ureh} \ln\left(1 - \chi_*^p
  \right) \\
  & - \dfrac{3 \bar{w}_\ureh+1}{3 + 3 \bar{w}_\ureh} 
\ln\left(\chi_*^{p-1} \right) \\
  & = \dfrac{\mu^2}{2p \Mp^2}\left(\chi_\uend^2 + \dfrac{2}{p-2}
    \chi_\uend^{2-p} \right) + \calFnuc,
\end{aligned}
\end{equation}
with
\begin{equation}
\begin{aligned}
  & \calFnuc = -N_0 + \dfrac{1+3\bar{w}_\ureh}{3+3\bar{w}_\ureh}\ln\left(2\pi
    \sqrt{120} \dfrac{\Qrms}{T} \right) \\ & - \dfrac{1}{3+3\bar{w}_\ureh}
  \ln\left[9 \left(\dfrac{\mu}{\Mp p} \right)^{3\bar{w}_\ureh+1}
    2^{(3\bar{w}_\ureh-1)/2}\right] \\ & - \dfrac{1}{3 +3 \bar{w}_\ureh} \ln
  \left(1-\chi_\uend^p \right) + \dfrac{1-3\bar{w}_\ureh}{3+3\bar{w}_\ureh} \ln
  \left(\dfrac{\rhonuc^{1/4}}{\Mp} \right).
\end{aligned}
\end{equation}
Similarly, solving for $\rhoreh=\rhoend$ gives $\chi_*^\uend$ as the
solution of
\begin{equation}
\label{eq:chistarend}
\begin{aligned}
 & \dfrac{\mu^2}{2p \Mp^2} \left(\chi_*^2 + \dfrac{2}{p-2} \chi_*^{2-p} \right)
  + \dfrac{1}{4} \ln(1-\chi_*^p) - \dfrac{1}{2} \ln(
  \chi_*^{p-1}) \\ & = \dfrac{\mu^2}{2p \Mp^2}\left(\chi_\uend^2 + \dfrac{2}{p-2}
    \chi_\uend^{2-p} \right) + \calFend,
\end{aligned}
\end{equation}
where 
\begin{equation}
\begin{aligned}
  \calFend = -N_0 & + \dfrac{1}{2}\ln\left(2\pi \sqrt{120}
    \dfrac{\Qrms}{T} \right) \\ & - \dfrac{1}{4} \ln\left[9
    \left(\dfrac{\mu}{\Mp p} \right)^{2} \left(1-\chi_\uend^p
    \right) \right].
\end{aligned}
\end{equation}
As a result, for given values of $\mu$, $p$ and $\bar{w}_\ureh$, one
has first to solve Eq.~(\ref{eq:chiend}) to get $\chi_\uend$, then
Eqs.~(\ref{eq:chistarnuc}) and (\ref{eq:chistarend}) to obtain
$\chi_*^\unuc$ and $\chi_*^\uend$ from which $\Delta N_*^\unuc$ and
$\Delta N_*^\uend$ are deduced by using Eq.~(\ref{eq:sftraj}). From
the value of $\chi_*$, one can also directly evaluate the two
slow-roll parameters $\epsilon_{1*}$ and $\epsilon_{2*}$. Let us
notice that some of the expressions above can be ill defined if
$p=2$. In this case, Eqs.~(\ref{eq:chistarnuc}) and
(\ref{eq:chistarend}) should be rederived from the start and one can
show that it then always leads to well-defined expressions. The rest
is the same as for the large field models (see
Sec.~\ref{subsec:largefield}), $\rhoreh$ being in one-to-one
correspondence with the value of $\Delta N_*$ through
Eqs.~(\ref{eq:Rrad}) and (\ref{eq:Rradw}) once a value of
$\bar{w}_{\ureh}$ has been chosen. It is worth emphasizing again that,
in order to derive these results, no assumption has been made about
the reheating epoch which is entirely characterized by $\rhoreh$ and
$\bar{w}_{\ureh}$.

\begin{figure*}
\begin{center}
\includegraphics[width=0.45\textwidth,clip=true]{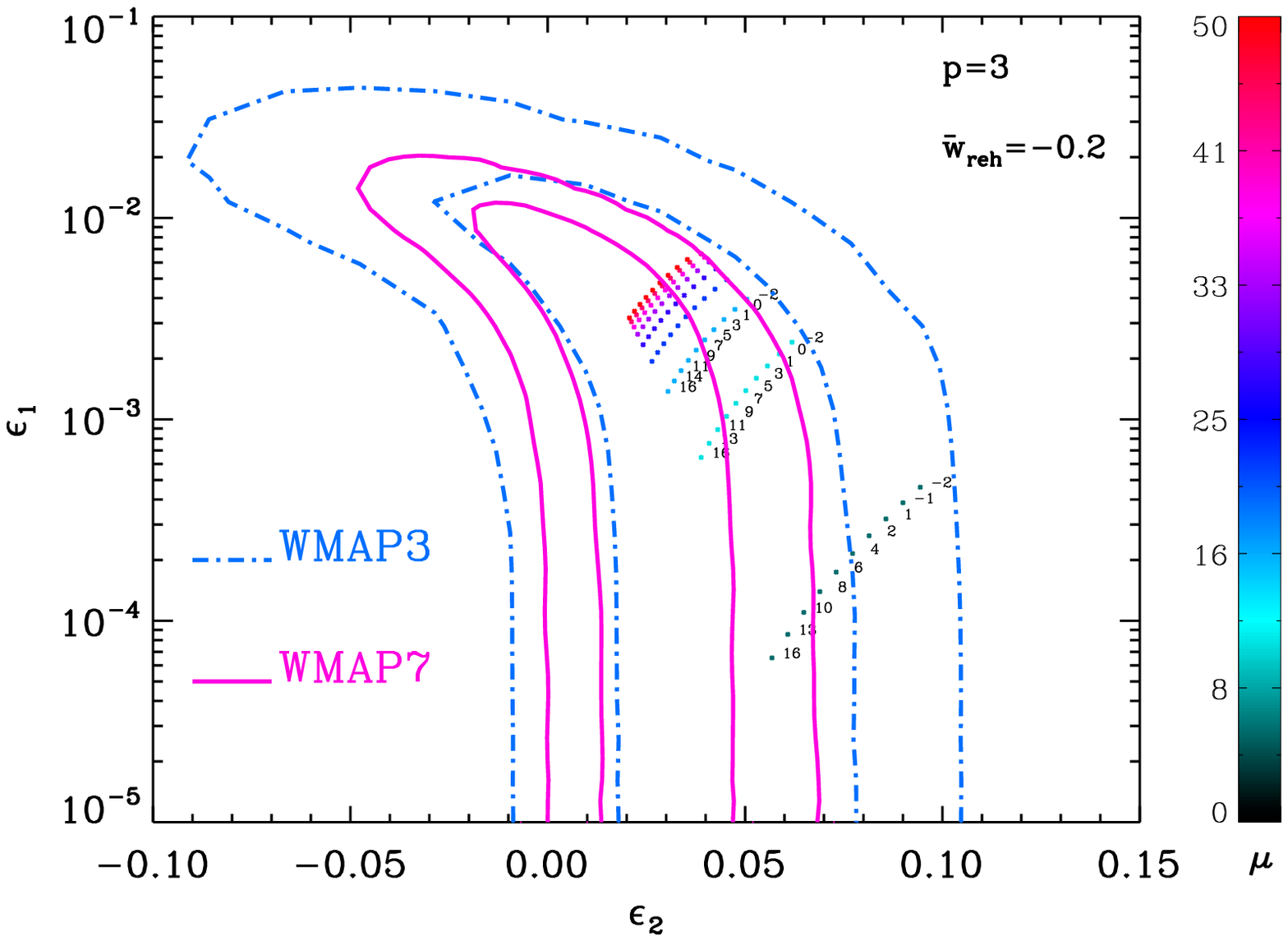}
\includegraphics[width=0.45\textwidth,clip=true]{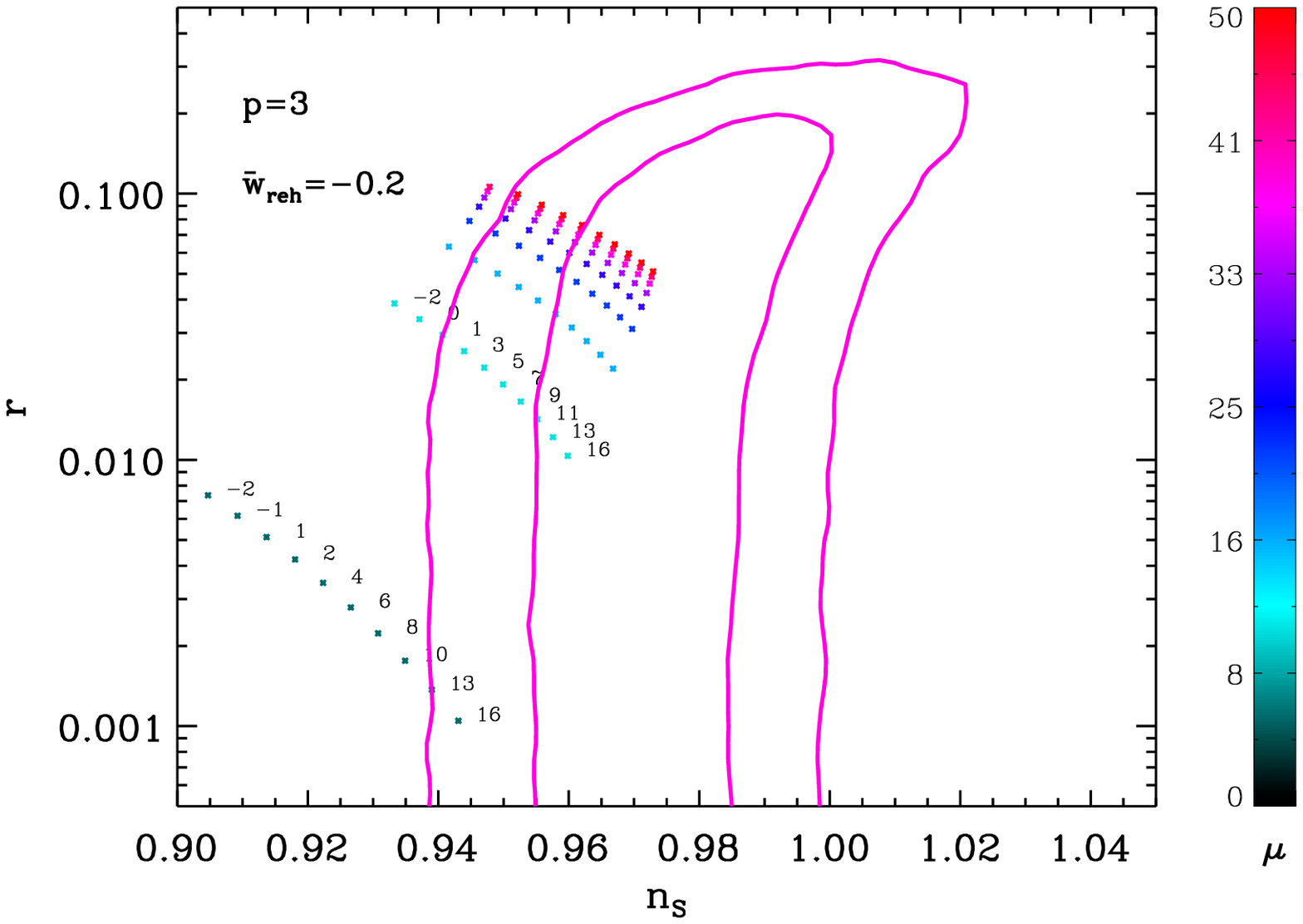}
\includegraphics[width=0.45\textwidth,clip=true]{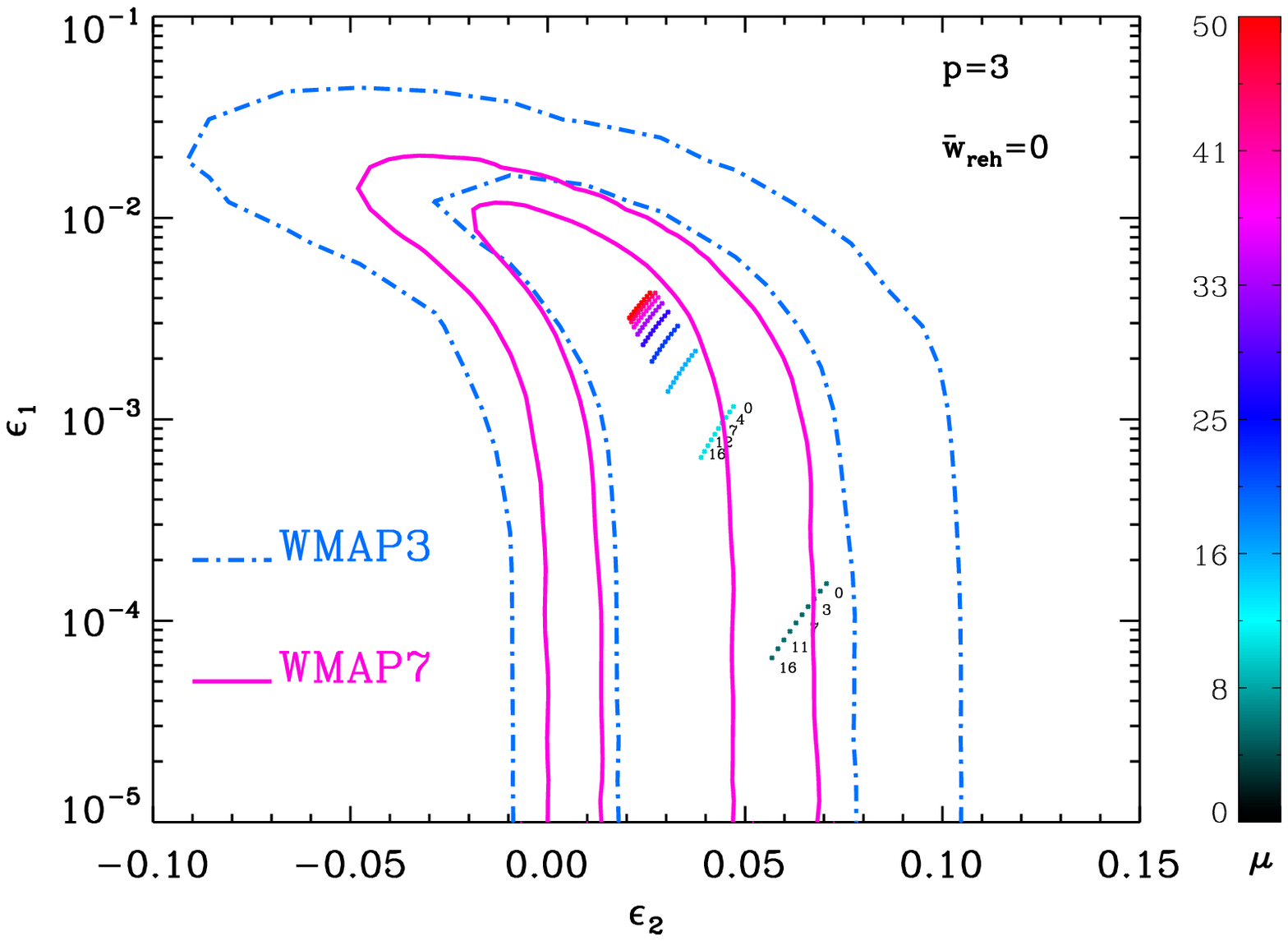}
\includegraphics[width=0.45\textwidth,clip=true]{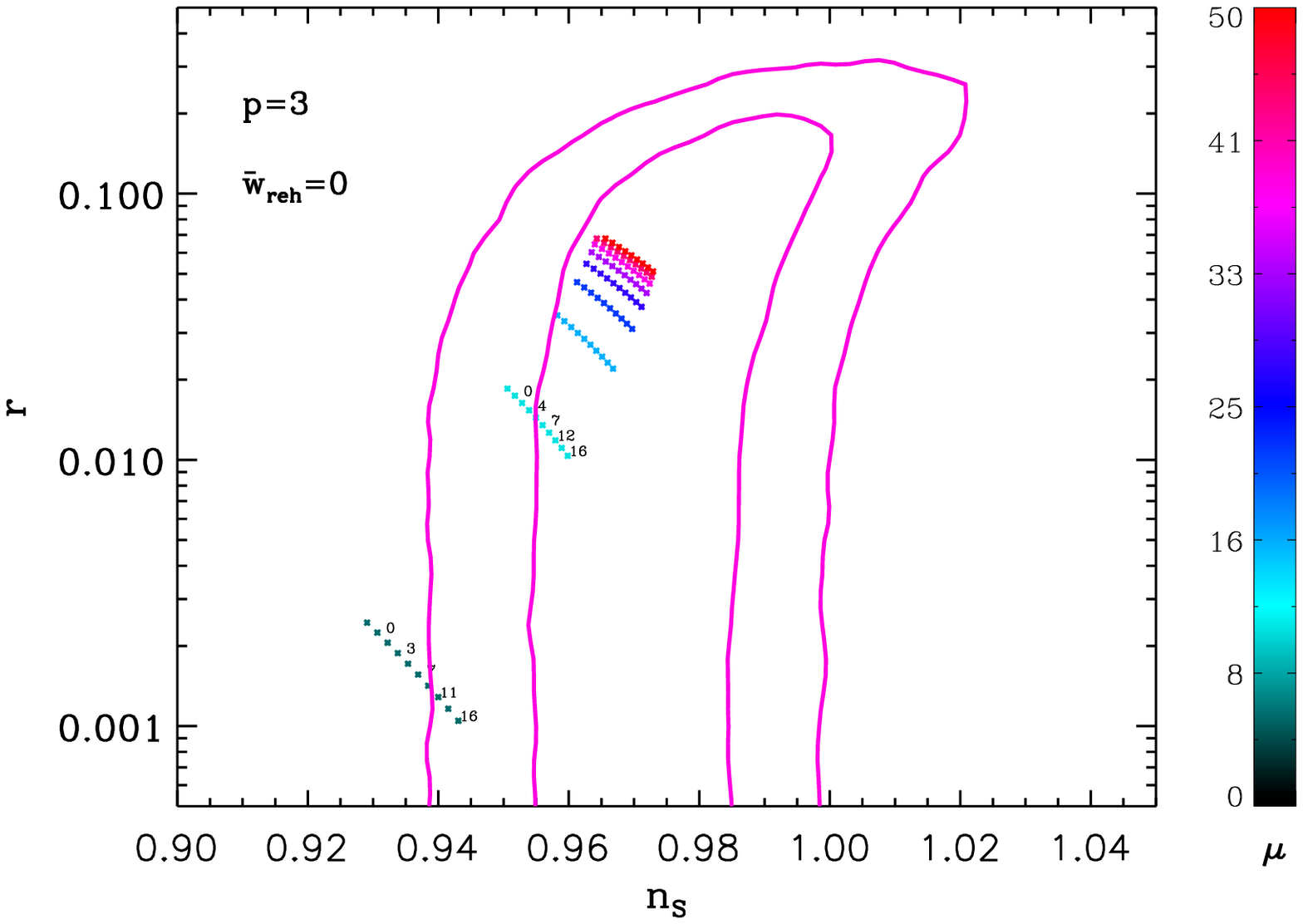}
\includegraphics[width=0.45\textwidth,clip=true]{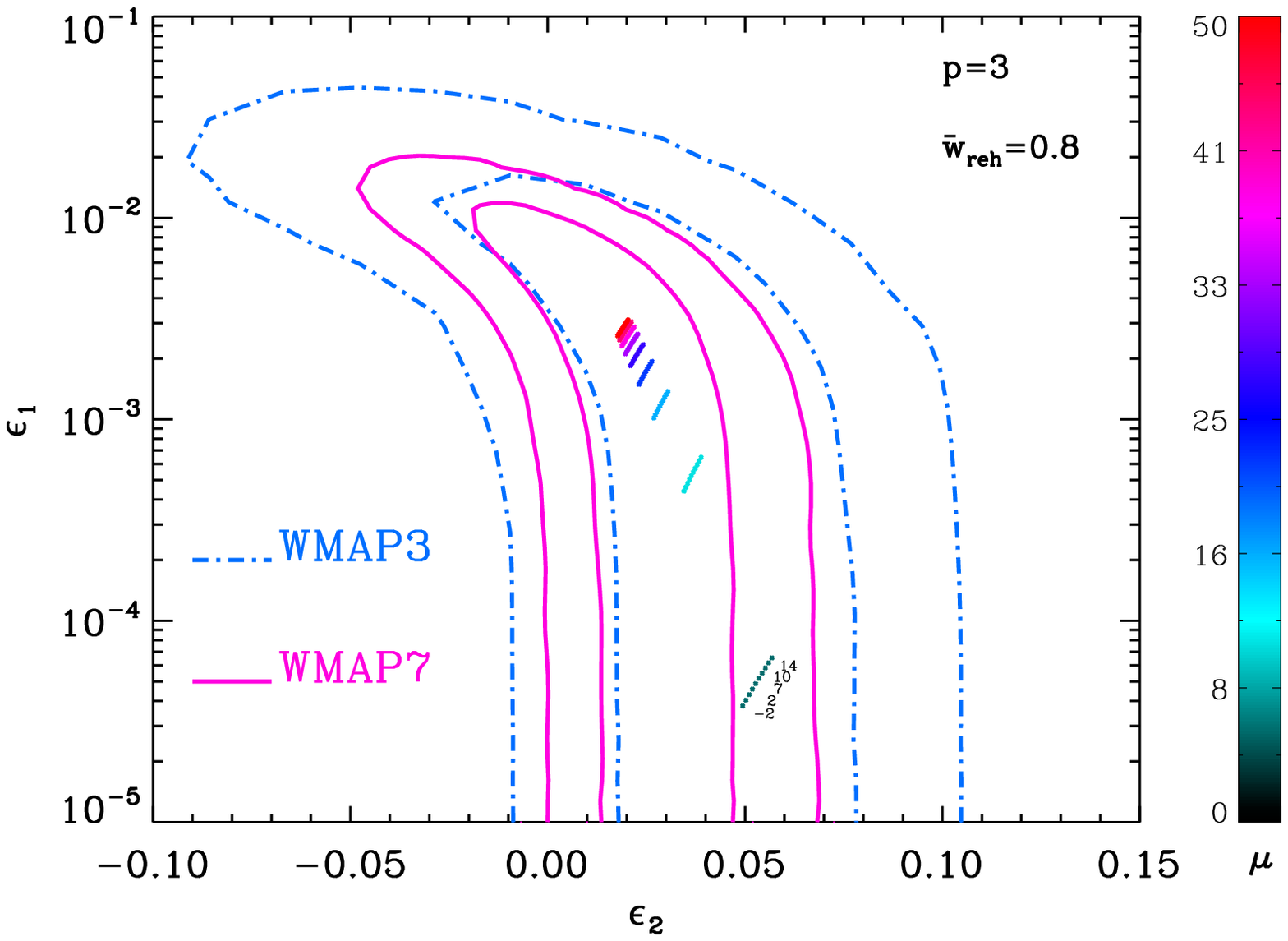}
\includegraphics[width=0.45\textwidth,clip=true]{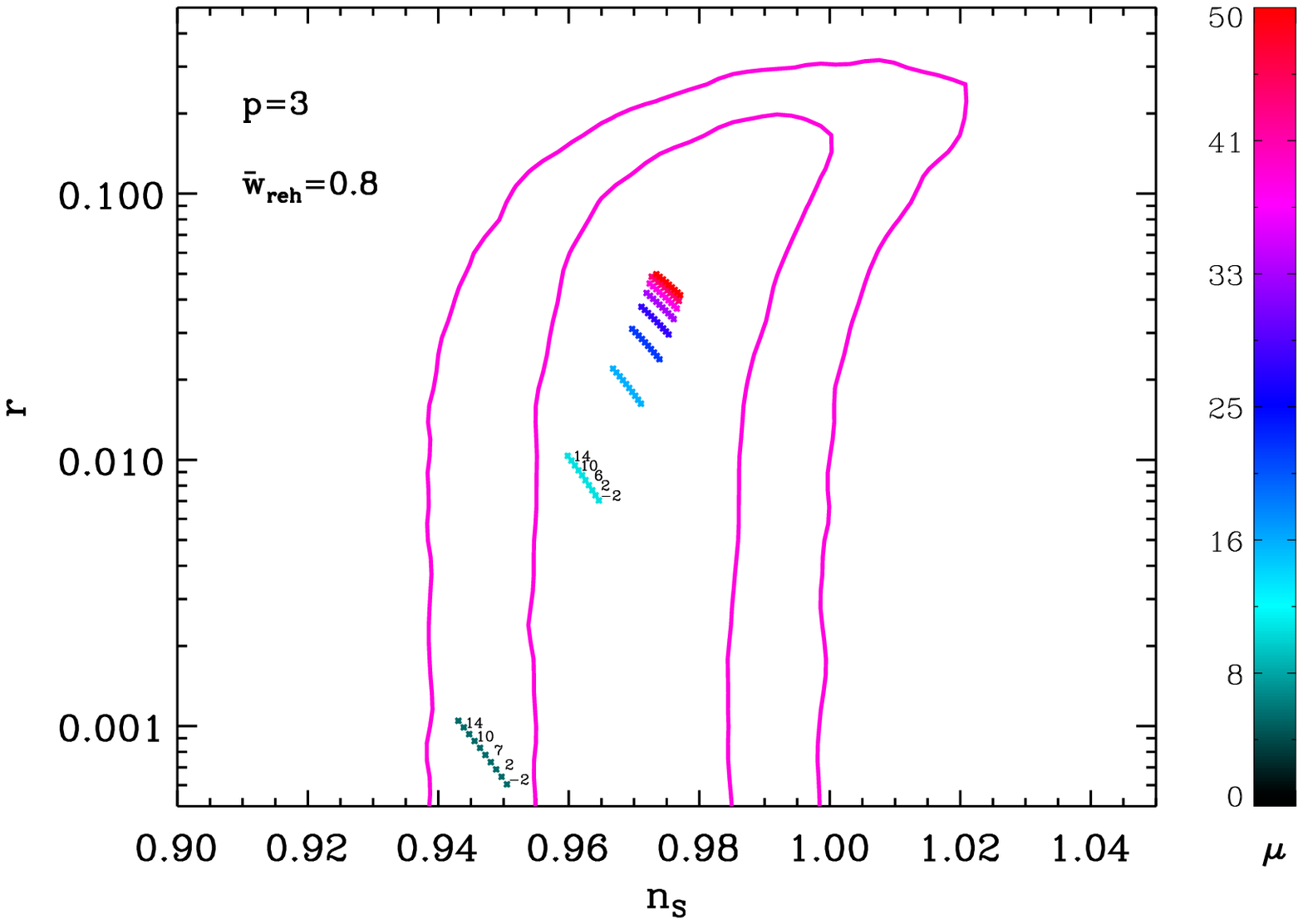}
\caption{Reheating consistent slow-roll predictions for the small
  field models with an assumed generic value $p=3$ and for
  $\bar{w}_\ureh=-0.2$ (top), $\bar{w}_\ureh=0$ (middle) and
  $\bar{w}_\ureh=0.8$ (bottom). The right panels display the
  corresponding predictions in the plane $(\nS,r)$. The reheating has
  a strong influence for low values of both $\bar{w}_\ureh$ and
  $\mu/\Mp$. As in Figs.~\ref{fig:srlfnsR} and~\ref{fig:srlf}, the
  annotations give the values of $\log(g_*^{1/4}T_{\ureh}/\GeV)$.
  It is also interesting to notice that the sequence of successive
  $T_\ureh$ is switched for $\bar{w}_\ureh=0.8$ (in fact for
  $\bar{w}_\ureh>1/3$), \ie large reheating temperatures correspond to
  smaller spectral indices for $\bar{w}_\ureh=0.8$ (\ie
  $\bar{w}_\ureh>1/3$) while, for $\bar{w}_\ureh=0$ or
  $\bar{w}_\ureh=-0.2$ (\ie $\bar{w}_\ureh<1/3$), they correspond to
  larger $\nS$.}
\label{fig:srsf}
\end{center}
\end{figure*}

In Fig.~\ref{fig:srsf}, we have represented the slow-roll predictions
for an assumed generic value of $p=3$ and various values of
$\bar{w}_\ureh$ ranging from $-0.2$ to $0.8$. The annotations are the
values of $\log(g_*^{1/4}T_{\ureh}/\GeV)$ while the color scale traces
the values of $\mu/\Mp$. For small field models, the reheating energy
scale is all the more so constrained that $\bar{w}_\ureh$ and $\mu$
are small. In fact, these plots show that $\bar{w}_\ureh$ and $\mu$
are degenerated: it is possible to render compatible a low value of
$\bar{w}_\ureh$ provided $\mu$ is super-Planckian. Conversely, small
values of $\mu$ can be made compatible with the data for a high energy
scale reheating if $\bar{w}_\ureh<1/3$, or a low energy scale
reheating for $\bar{w}_\ureh>1/3$. In the next section, we perform a
full analysis of the small field models, reheating included, in view
of the WMAP7 data to quantify the above claims in terms of posterior
probability distributions.

\section{Inferring reheating from CMB data}
\label{sec:cmbtoreh}

In view of the previous results, the correct way to discuss how well
the CMB data constrain a set of known inflationary models is to
perform a Bayesian analysis of the data given the model parameters,
including the reheating. Notice that this is different than
constraining the slow-roll parameters, or the spectral index and
tensor-to-scalar ratio, which only encode the shape of the primordial
power spectra and know nothing about reheating (whereas a model of
inflation does).

\subsection{Exact numerical integration}
\label{subsec:exact}

The numerical exact integration method has been introduced in
Refs.~\cite{Ringeval:2005yn, Martin:2006rs,Ringeval:2007am,
  Lorenz:2007ze} and consists of the computation of the primordial
power spectra assuming only General Relativity and linear perturbation
theory. Therefore, the only model parameters are the ones appearing in
the inflaton potential together with the reheating parameter $\Rrad$,
for the very reasons explained in Sec.~\ref{sec:physicalorigin}. The
numerical integration of the inflationary perturbations sets up the
initial conditions for the subsequent cosmological perturbations from
which the CMB anisotropies are deduced. For this purpose, we have used
a modified version of the \texttt{CAMB} code~\cite{Lewis:1999bs}
coupled to a Monte--Carlo--Markov--Chain (MCMC) exploration of the
parameter space implemented in the \texttt{COSMOMC}
code~\cite{Lewis:2002ah} and given the WMAP7
data~\cite{Komatsu:2010fb, Larson:2010gs, Jarosik:2010iu}. Concerning
the standard cosmological model, we have assumed a flat $\Lambda$CDM
model having five parameters: the density parameter of baryons,
$\OmegaB$, of cold dark matter $\OmegaDM$, the Hubble parameter today
$H_0$, the optical depth $\tau$ encoding the redshift at which the
Universe reionized, and the nuisance parameter $\Asz$ encoding the
relative amplitude of the diffuse Sunyaev--Zel'dovich (SZ) effect
compared to the analytical model of Ref.~\cite{Komatsu:2002wc}. In
fact, as discussed in Ref.~\cite{Lewis:2002ah}, it is more convenient
to sample the cosmological parameter space along the rescaled quantity
$(\OmegaB h^2, \OmegaDM h^2, \tau, \theta, \Asz)$ where $H_0 = 100 h
\, \km/\sec/\Mpc$ and $\theta$ measures the ratio of the sound horizon
at last scattering to the angular diameter distance. Following
Ref.~\cite{Komatsu:2010fb}, we have included the lensing corrections
on the temperature and polarization power spectra, and, to limit
parameter degeneracies, completed the WMAP7 data with the latest
Hubble Space Telescope (HST) bound on
$H_0$~\cite{Riess:2009pu}. Concerning the primordial parameters, they
are now provided by our inflationary model parameters, up to some
observationally convenient rescaling. For instance, we will prefer to
sample on $P_*$, the amplitude of the scalar perturbation at the pivot
scale, rather than on the potential normalization $M$, both being in
one-to-one correspondence. Similarly, it is more convenient to sample
the reheating era over the parameter $\Rreh$,
\begin{equation}
\label{eq:Rrehdef}
\Rreh \equiv \Rrad  \dfrac{\rhoend^{1/4}}{\Mp}\,,
\end{equation}
rather than $\Rrad$. As can be seen by comparing Eq.~(\ref{eq:Rradsr})
and the following exact expression:
\begin{equation}
\label{eq:Rrehsr}
\begin{aligned}
  \ln R = \Nend - \Nstar + \Nzero + \dfrac{1}{2} \ln\left(3
    \dfrac{V_\uend}{V_*} \dfrac{3 - \epsilon_{1*}}{3 -
      \epsilon_{1\uend}} \right), 
\end{aligned}
\end{equation}
contrary to $\Rreh$, the values of $\Rrad$ explicitly depend on
$P_*$. This would induce unwanted correlations between $\Rrad$ and
$P_*$ which are therefore avoided by sampling the reheating over
$\Rreh$. Notice that since $\Rreh$ and $\Rrad$ differ by a factor
$\rhoend$, they are also in one-to-one correspondence once the model
of inflation is specified~\cite{Martin:2006rs}. In order to perform
the MCMC analysis, we still have to specify the prior probability
distributions. Concerning the cosmological parameters, we have chosen
wide flat priors around the preferred posterior values obtained by the
WMAP team~\cite{Komatsu:2010fb}. The reheating energy scale being
unknown, we assume a flat prior on $\ln \Rreh$ whose extension is
given by the consistency conditions mentioned in
Sec.~\ref{subsec:why}. Reheating should occur before nucleosynthesis
and after the end of inflation while the positivity energy conditions
imply $-1/3< \bar{w}_\ureh<1$. As a result, we take a flat prior for
$\ln R$ in the range~\cite{Martin:2006rs}
\begin{equation}
  \ln\left(\dfrac{\rhonuc^{1/4}}{\Mp} \right) < 
\ln \Rreh < -\dfrac{1}{3} 
\ln\left(\dfrac{\rhonuc^{1/4}}{\Mp} \right) 
+ \dfrac{4}{3} \ln\left(\dfrac{\rhoend^{1/4}}{\Mp} \right).
\end{equation}
The lower bound is approximately $\simeq -47$ for $\rhonuc =10 \,\MeV$
whereas the upper bound depends on $\rhoend$, and thus on the other
inflationary model parameters. Finally, we have chosen a flat prior on
the logarithm of $P_*$ around the value giving the right amplitude of
the CMB anisotropies: $2.7 < \ln\left(10^{10} P_*\right) < 4.0$. Let
us notice that since $P_*$ is well constrained, this translates into
an upper bound on $\ln (\rhoend^{1/4}/\Mp) < -5.5$ (analogous to the
upper bound on $H_*$ in the slow-roll approximation) that will
therefore be inherited by $\ln \Rreh$ so that the maximal value of the
upper bound is $\simeq 8.3$. The other prior choices on the primordial
parameters are those concerning the inflaton potential and will be
specified later.

\par

In the following, we perform the WMAP7 data analysis along those lines
for both the large and small field models. As the first step, we sample
over the rescaled reheating parameter $\ln \Rreh$ without any
assumptions on $\bar{w}_\ureh$. We show that it is actually
constrained for all models. As can be checked in Eq.~(\ref{eq:Rrad}),
it means that the CMB data restrict the \emph{a priori} possible values of
$\Delta N$ and $\bar{w}_\ureh$. Conversely, this result shows that not
including the reheating parameter when constraining inflationary
models is no longer a reasonable option. For the second step, we break
the degeneracy between $\bar{w}_\ureh$ and $\Delta N$ and assume that
$\bar{w}_\ureh$ takes its natural values for the large field models
given in Eq.~(\ref{eq:wrehlf}), or choose a specific value in the
small field models. These reasonable extra assumptions translate the
bounds on $\ln \Rreh$ into a lower limit on $\rhoreh$ and/or
$g_*^{1/4}T_{\ureh}$. Unless specified, we have stopped the MCMC
exploration according to the R--statistics~\cite{Gelman:1992}
implemented in \texttt{COSMOMC} such that the difference in variances
between the different Markov chains does not exceed a few
percent. Typically, this corresponds to a set of $300\,000$ to
$500\,000$ samples depending on the underlying model of inflation.
\begin{figure}
\begin{center}
\includegraphics[width=8.9cm]{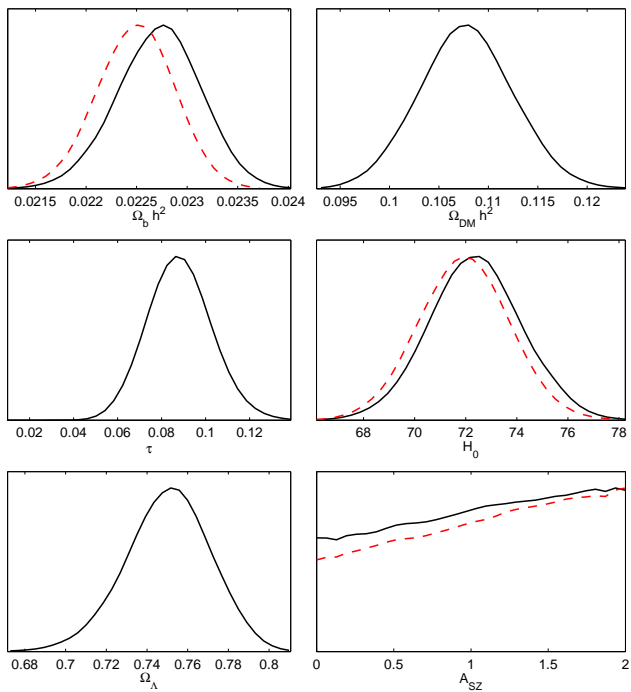}
\caption{Marginalized posterior probability distributions for the base
  and derived cosmological parameters in large field inflation. The
  black solid lines are without any assumptions on the large field
  reheating whereas the red dashed ones are under the prior
  $\bar{w}_\ureh=(p-2)/(p+2)$. They have been represented only when they
  differ with respect to the former.}
\label{fig:lfcosmo}
\end{center}
\end{figure}

\subsection{Large field models}
\label{subsec:lfnumerics}

The potential for the large field models is given in
Eq.~(\ref{eq:lfpot}). Together with $P_*$ and $\ln\Rreh$, there is
only one additional primordial parameter $p$ for which we have chosen
a flat prior in the range $p\in[0.2,5]$. The upper bound is motivated
by the previous constraints on large fields~\cite{Martin:2006rs}
whereas the lower one is a theoretical prejudice associated with the
non-naturalness of extremely small values of $p$ in any field
theory. The marginalized posterior distributions for the sampled and
derived cosmological parameters are represented in
Fig.~\ref{fig:lfcosmo} for the two prior assumptions detailed in the
following. The solid lines are without any assumption on the reheating
whereas the dashed ones are under the natural equation of state
$\bar{w}_\ureh=(p-2)/(p+2)$. These probabilities are compatible with
the one already derived in the literature~\cite{Lorenz:2008je,
  Komatsu:2010fb} up to slight shifts coming from changing the
reheating assumptions. This is the result of some tension between the
large field models which generically predict a large tensor-to-scalar
ratio $r$ and its non-observation. Being more restrictive on the
reheating gives less flexibility to the model such that $H_0$ and
$\OmegaB$ are slightly shifted to compensate for the too high $r$
values.
\begin{figure}
\begin{center}
\includegraphics[width=9cm]{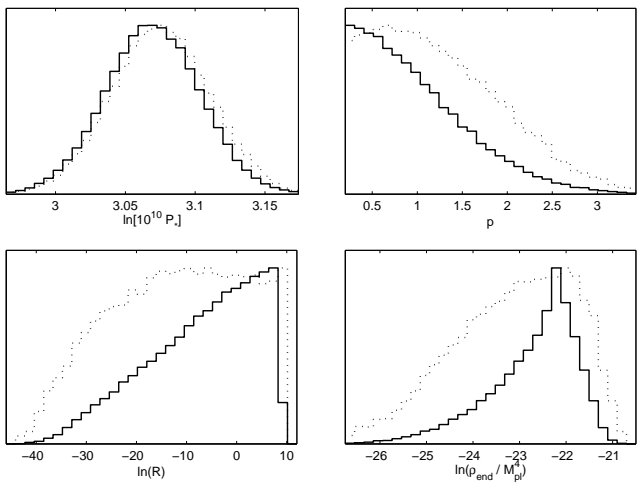}
\caption{Marginalized posterior probability distributions (solid
  lines) and mean likelihoods (dotted) for the large field model
  primordial parameters. This is without assumption on the reheating
  era. Notice the lower bound on the reheating parameter $\ln R$ which
  correlates with the potential power $p$ (see also
  Fig.~\ref{fig:lf2D}). The energy scale at the end of large field
  inflation is also constrained.}
\label{fig:lf1D}
\end{center}
\end{figure}

\begin{figure}
\begin{center}
\includegraphics[width=9cm]{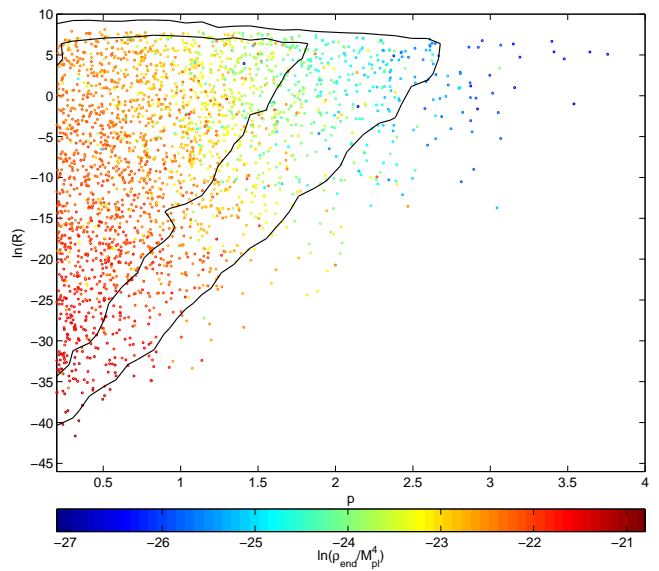}
\caption{Two-dimensional marginalized posterior probability
  distribution (point density) in the plane $(p, \ln \Rreh)$ and its
  one- and two-sigma confidence intervals. Correlations with the
  energy scale of large field inflation are traced by the color
  scale.}
\label{fig:lf2D}
\end{center}
\end{figure}

In Fig.~\ref{fig:lf1D}, we have plotted the marginalized probability
distribution for the large field primordial parameters without
assumption on the reheating. It is particularly interesting to compare
these plots to Fig.~18 of Ref.~\cite{Martin:2006rs} since this allows
us to see the improvements on the parameter constraints coming from
the passage from WMAP3 data to WMAP7. In addition to the expected
constraints on $P_*$, we find the $95\%$ confidence limit
\begin{equation}
\label{eq:lfpmax}
p < 2.2\,,
\end{equation}
suggesting that $\phi^2$ inflation may now be considered under
pressure. Let us emphasize that this result is robust against any
possible reheating evolution since marginalized over $\ln
\Rreh$. Concerning this last parameter, we find a $95\%$ lower bound:
\begin{equation}
\label{eq:lfRmin}
\ln \Rreh > -28.9\,.
\end{equation}
In fact, as can be checked in Fig.~\ref{fig:lf2D}, these two
parameters are correlated together and also with $\rhoend$. These
correlations can be understood as follows. From Eq.~(\ref{eq:Rrehsr}),
the quantities $\ln R$, $p$ and $\ln\left(\rhoend/\Mp^4\right)$ are
related by the formula
\begin{eqnarray}
\ln \left(\frac{\rhoend}{\Mp^4}\right)&=&\ln \left(128\pi^2P_*\right)-2N_0
+2\ln R-\frac{2}{1-\nS}\nonumber \\ & & -\frac{p}{2}\frac{1+\nS}{1-\nS}+
\ln\left[\frac{8p(1-\nS)}{p+2}\right],
\end{eqnarray}

\noindent where $P_*$ and $\nS$ are well constrained quantities. As a
result, at fixed $\ln R$, the larger the $p$ values, the lower the
energy scale at the end of inflation has to be, which is exactly what
is observed in Fig.~\ref{fig:lf2D}. Of course, $p$ cannot be too large
since, in this case, the tensor-to-scalar ratio $r$ increases and
rapidly becomes incompatible with the CMB data. In
Fig.~\ref{fig:lf2D}, we also observe that the smaller $p$, the larger
the allowed range of variation of $\ln R$. The upper limit on $\ln R$
does not depend on $p$ and just comes from the upper limit on the
energy scale of inflation ($\ln R \lesssim 8.3$). On the other hand,
the lower limit strongly depends on $p$ and represents a non trivial
result. This expresses the fact that, for a given $p$, there are
values of $\ln R$ for which there is no way to obtain, at the same
time, a consistent reheating epoch and CMB predictions compatible with
the data. From this effect, we also get the energy scale of large
field inflation, and at two-sigma level
\begin{equation}
\label{eq:rhoendbound}
4.4 \times 10^{15} \GeV < \rhoend^{1/4} < 1.2 \times 10^{16} \GeV\,.
\end{equation}
The upper limit just comes from the constrain on the energy scale of
inflation while the lower limit originates from the fact that $p$
cannot be too large (recall low values of $\rhoend$ means large value
of $p$).

\begin{figure}
\begin{center}
\includegraphics[width=8.9cm]{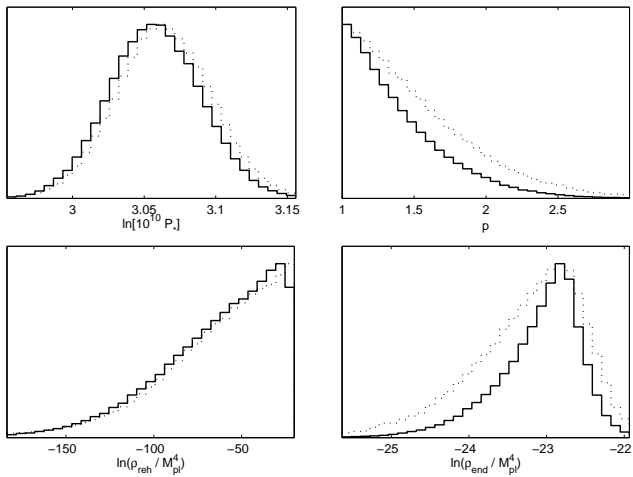}
\caption{Marginalized posterior probability distributions (solid lines)
  and mean likelihood (dotted lines) for the large field parameters when
  $\bar{w}_\ureh=(p-2)/(p+2)$. This extra-assumption on reheating
  yields to tighter constraints than in Fig.~\ref{fig:lf1D}. In
  particular, we find $\rhoreh > 17.3\, \TeV$ at $95\%$ confidence
  level, as well as $p<2.1$.}
\label{fig:lfwreh1D}
\end{center}
\end{figure}
\begin{figure}
\begin{center}
\includegraphics[width=9cm]{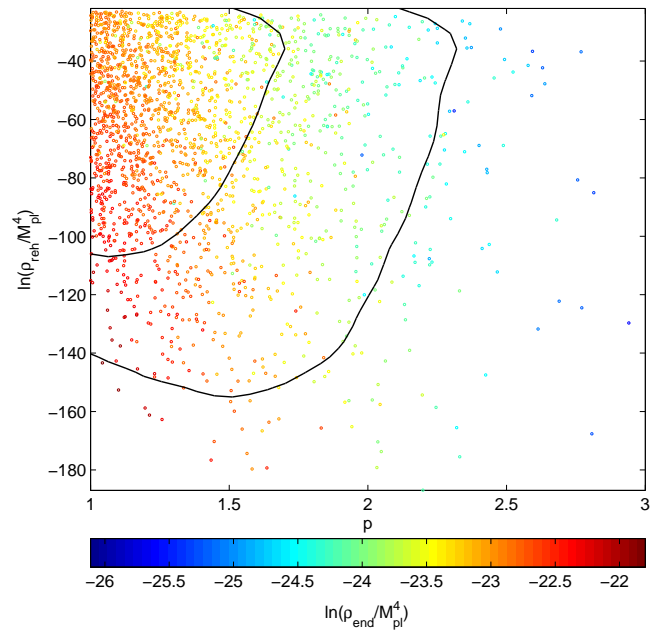}
\caption{One- and two-sigma marginalized limits in the plane
  $[p,\ln(\rhoreh/\Mp^4)]$ for the large field models with
  $\bar{w}_\ureh=(p-2)/(p+2)$. The point density traces the associated
  two-dimensional posterior while the color map shows correlations
  with the energy scale at which inflation ends.}
\label{fig:lfwreh2D}
\end{center}
\end{figure}

Assuming now that the reheating proceeds according to
Eq.~(\ref{eq:wrehlf}), one obtains the marginalized posteriors plotted
in Fig.~\ref{fig:lfwreh1D} and Fig.~\ref{fig:lfwreh2D}. To be
consistent, we have modified our prior on $p$ by assuming a flat
distribution in $p\in]1,5]$; the case $p=1$ is a limiting case that
may be problematic. Indeed, values of $p<1$ would induce
$\bar{w}_\ureh<-1/3$ and inflation would not stop. Moreover, instead
of using $\ln \Rreh$, we have used Eqs.~(\ref{eq:Rradw}) and
(\ref{eq:Rrehdef}) to sample the parameter space over $\ln
(\rhoreh/\Mp^4)$ and from a flat prior in $[\ln(\rhonuc/\Mp^4),
\ln(\rhoend/\Mp^4)]$. The upper bound on the $p$ posterior is slightly
tighter due to the restriction made over the reheating: we find the
two-sigma limit $p<2.1$. For the same reasons, the energy scale of
large field inflation is a bit more constrained, and the two-sigma
range becomes
\begin{equation}
  5.2 \times 10^{15}\,\GeV < \rhoend^{1/4} < 9.1\times 10^{15}\, \GeV.
\end{equation}
Certainly, the more interesting result is the lower bound on the
reheating energy scale. At $95\%$ of the confidence limit
\begin{equation}
\label{eq:lfrhorehmin}
\rhoreh^{1/4} > 17.3\, \TeV\,.
\end{equation}
The correlations between these three parameters are represented in
Fig.~\ref{fig:lfwreh2D} and have the same origin as the ones
displayed in Fig.~\ref{fig:lf2D}, up to the change of variable $\Rreh$
to $\rhoreh$. In particular, we see that, at a fixed value of $p$, the
constraints on $\rhoreh$ are tighter for $p \lesssim 1.5$ than for $p
\simeq 1.5$. This comes from the fact that the reheating is well
constrained for a negative mean equation of state, which precisely
corresponds to $p<2$ [see Eq.~(\ref{eq:wrehlf})]. The change of
behavior around $p=1.5$ comes from this effect combined with a two
high tensor-to-scalar ratio when $p\gtrsim 2$. Let us also emphasize
that Eq.~(\ref{eq:lfrhorehmin}) is marginalized over all large field
models. Coming with a theoretical preference for a given value of $p$
can lead to stronger bounds, as for instance if $p \gtrsim 1$ or $p=2$ (see
Fig.~\ref{fig:lfwreh2D}). Finally, one can check that the bounds found
in this section are compatible with the expectations we have derived
from the slow-roll predictions of Sec.~\ref{subsec:largefield}.

\par

In the next section, we perform a similar analysis for the small field
models.

\subsection{Small field models}
\label{subsec:sfnumerics}

\begin{figure}
\begin{center}
\includegraphics[width=9cm]{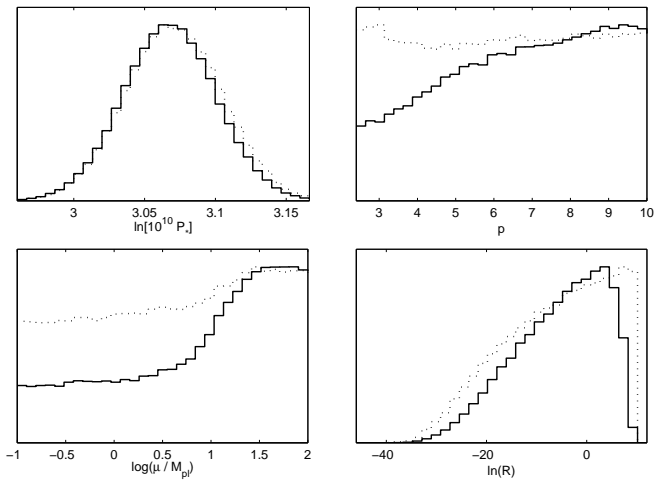}
\caption{Marginalized posterior probability distributions (solid
  lines) and mean likelihoods (dotted) for the small field model
  primordial parameters. This is without assumption on the reheating
  era. Notice again the lower bound on the reheating parameter $\ln R$
  and the slightly favored super-Planckian values of
  $\mu$. Correlations are displayed in Fig.~\ref{fig:sf2D}.}
\label{fig:sf1D}
\end{center}
\end{figure}

\begin{figure*}
\begin{center}
\includegraphics[width=18cm]{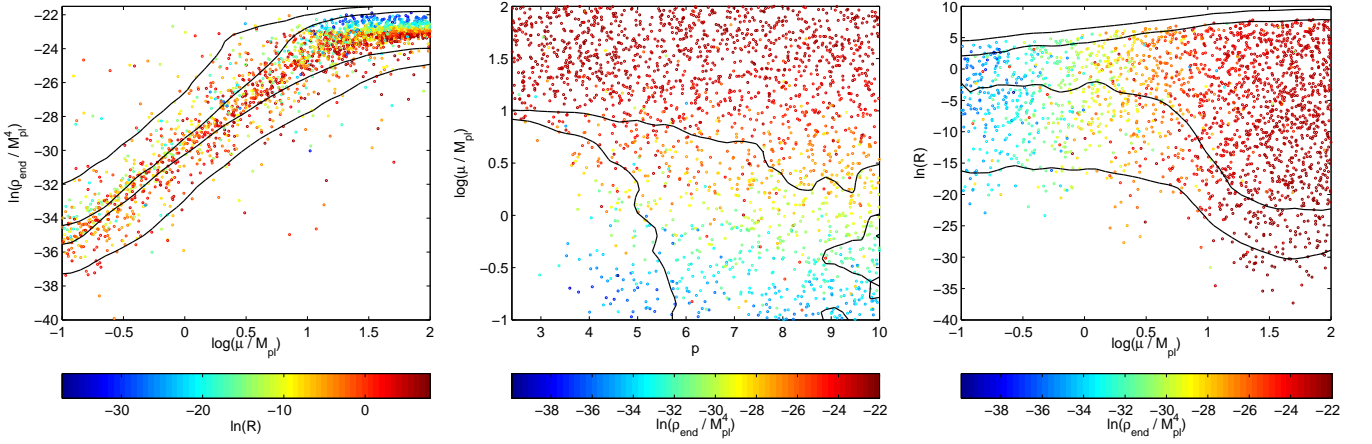}
\caption{One- and two-sigma contours of the two-dimensional
  marginalized probability distributions for small field
  inflation. The point density traces the associated two-dimensional
  posterior while the color scale shows correlations with third
  parameters. Notice the correlation between $\mu$ and $p$, as well as
  with $\ln \Rreh$.}
\label{fig:sf2D}
\end{center}
\end{figure*}

The small field model potential of Eq.~(\ref{eq:sfpot}) involves an
extra parameter compared to large fields which is the VEV $\mu$. The
scale of this parameter being unknown, we have chosen a flat prior on
$\log(\mu/\Mp)$ in the range $[-1,2]$. With the lower and upper limits
being chosen only for numerical convenience, one should keep in mind
that the physical values of $\mu$ may be larger or smaller. The
important point is however that such a prior includes both
sub-Planckian and super-Planckian values without prejudice. Concerning
the potential index $p$, we have chosen a flat prior in the range
$p\in[2.4,10]$. The upper bound is arbitrary whereas the lower bound
excludes $p=2$ since this model is a special
case~\cite{Martin:2006rs}. In fact, there exists another reason that
justifies the above choices. In the limit $\mu/\Mp\gg 1$, one can
show, using a perturbative expansion in $\Mp/\mu$, that the two first
horizon flow functions $\epsilon_1$ and $\epsilon_2$ become
independent from $\mu $ and $p$, namely $\epsilon_1=(4\Delta
N_*+1)^{-1}$ and $\epsilon_2=4\epsilon_1$. Therefore, it would be
useless to take a larger upper bound on the $\mu$ prior since the
corresponding physical predictions are no longer affected by this
choice. All the other priors, both on the cosmological and primordial
parameters, have been chosen as for the large field exploration. For
the sake of clarity, we have not represented the marginalized
posteriors for the cosmological parameters. Contrary to the large
field models, these posteriors are almost the same whatever we assume
for the reheating. The reason is that small field models do not
have a tendency to produce a high tensor-to-scalar ratio. There is
therefore no need for the cosmological parameters to compensate for
such an effect and they decouple from the details of the inflationary
and reheating phases. Finally, the cosmological parameter posteriors in
small field inflation end up being very similar to the dashed curves
(or solid curves when they are absent) plotted in Fig.~\ref{fig:lf1D}.

\begin{figure}
\begin{center}
\includegraphics[width=8.8cm]{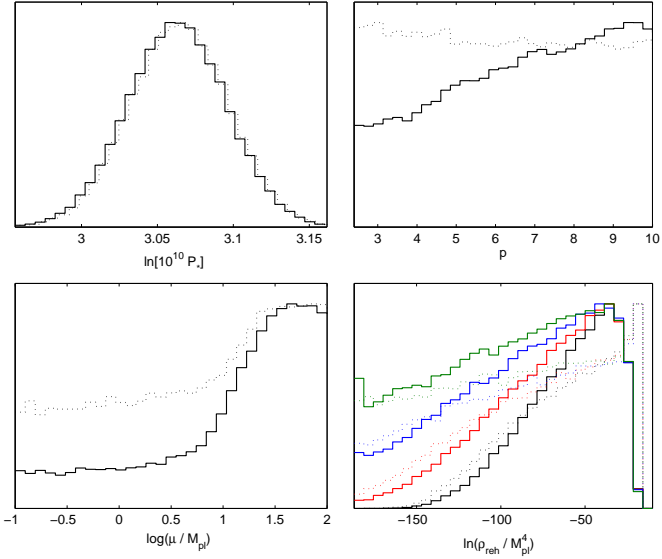}
\caption{Marginalized posterior probability distributions (solid
  lines) and mean likelihoods (dotted lines) for the small field
  model. These posteriors assume a constant equation of state with
  (from bottom to top) $\bar{w}_\ureh=-0.3$ (black lines), $-0.2$ (red
  lines), $-0.1$ (blue lines) and $\bar{w}_\ureh=0$ (green
  lines). Only $\rhoreh$ is affected by such prior choices (bottom
  right panel). The energy scale of reheating is all the more
  constrained from below than $\bar{w}_\ureh$ is small.}
\label{fig:sfwreh1D}
\end{center}
\end{figure}

In Fig.~\ref{fig:sf1D} and Fig.~\ref{fig:sf2D}, we have plotted the
marginalized posterior probability distributions (one- and
two-dimensional, respectively) for the primordial small field
parameters without assumption on the reheating. Again, we find the
WMAP data to give a lower limit on the reheating parameter, at
two-sigma level
\begin{equation}
\label{eq:sfRrehmin}
\ln \Rreh > -23.1\,.
\end{equation}
Concerning the parameters $\mu$ and $p$, they are not
constrained. However, the posteriors of Fig.~\ref{fig:sf1D} clearly
show a tendency to favor super-Planckian values of $\mu$ together
with large values of $p$. As can be seen in Fig.~\ref{fig:sf2D}, since
$\mu$ can take arbitrarily low values, the energy scale of small field
inflation is not constrained from below. We find only the consistency
condition that $\rhoend^{1/4} < 9 \times 10^{15}\, \GeV$ from the
$P_*$ limits. The correlations between $\mu$ and $p$ can be understood
from Sec.~\ref{subsec:smallfield} and come from the requirements of
having the right spectral index. The bound of Eq.~(\ref{eq:sfRrehmin})
has the same origin but through the selection of the favored $\Delta
N_*$ values. More details on these effects can be found in
Refs.~\cite{Martin:2006rs, Ringeval:2007am}.

\begin{figure}
\begin{center}
\includegraphics[width=9cm]{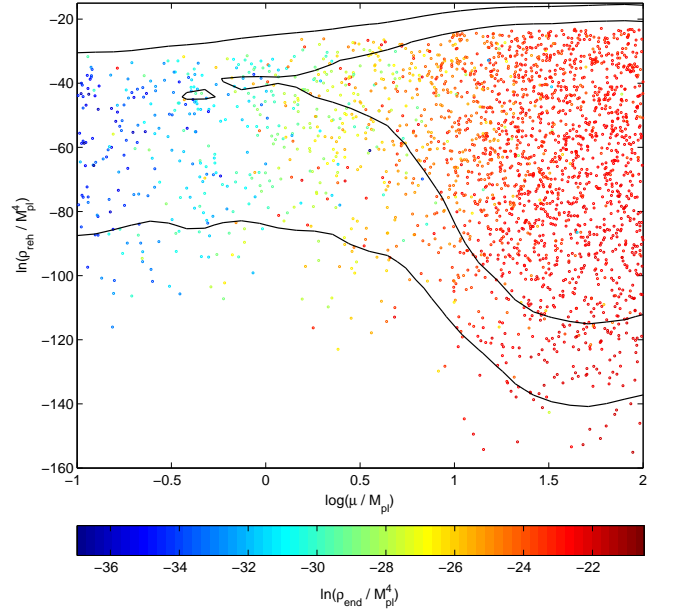}
\caption{One and two-sigma marginalized limits in the plane $[\log(\mu
  / \Mp),\ln(\rhoreh/\Mp^4)]$ for the large field models with
  $\bar{w}_\ureh=-0.3$. The point density traces the associated
  two-dimensional posterior while the color map shows correlations
  with the energy scale at which inflation ends.}
\label{fig:sfwreh2D}
\end{center}
\end{figure}

As we did for large fields, we now assume an equation of state
parameter for the small field reheating. Contrary to large field
models, we no longer have a relationship between $\bar{w}_\ureh$ and
the inflationary potential and one may only assume fiducial values
ranging from $-1/3$ to $1$. Again, the MCMC has been now sampled
directly over $\ln(\rhoreh/\Mp^4)$ rather than over $\ln \Rreh$ by
making use of Eqs.~(\ref{eq:Rradw}) and (\ref{eq:Rrehdef}). The
resulting one-dimensional posteriors for the primordial parameters
have been plotted in Fig.~\ref{fig:sfwreh1D} for the four values
$\bar{w}_\ureh=-0.3$, $-0.2$, $-0.1$ and $\bar{w}_\ureh=0$, and only
when the posteriors are affected by this choice. Comparing
Figs.~\ref{fig:sf1D} and Fig.~\ref{fig:sfwreh1D} shows that the
posteriors of $\mu$, $p$ (and $P_*$) are mostly independent of the
details of the reheating. The only quantity changing accordingly is
$\rhoreh$. This is not surprising since at a given $\Rreh$, changing
the values of $\bar{w}_\ureh$ modifies the number of e-folds the
Universe reheated. As a result, the constraint on $\Rreh$ for small
fields translates into a lower bound on $\rhoreh$ but only when
$\bar{w}_\ureh$ is small. At $95\%$ of confidence, we find
\begin{equation}
\begin{aligned}
\label{eq:conssf}
\bar{w}_\ureh & = -0.3 \quad \Rightarrow \quad 
\rhoreh^{1/4} > 8.9 \times 10^5\,\GeV,\\
\bar{w}_\ureh & = -0.2 \quad \Rightarrow \quad 
\rhoreh^{1/4} > 3.9 \times 10^2\, \GeV,
\end{aligned}
\end{equation}
while higher values of $\bar{w}_\ureh$ do not constrain $\rhoreh$ more
than the prior $\rhoreh>\rhonuc$. Physically, these results can be
fully understood from Fig.~\ref{fig:srsf}. In particular, the fact
that constraints on $\rhoreh$ can be derived for negative values of
$\bar{w}_\ureh$ is apparent from these plots. As expected,
correlations between $\rhoreh$ and the other parameters are similar to
the ones associated with $\ln \Rreh$. In Fig.~\ref{fig:sfwreh2D}, we
have represented the two-dimensional posterior and its one- and
two-sigma contours in the plane
$[\log(\mu/\Mp),\ln(\rhoreh/\Mp^4)]$. The color scale traces
correlations with the energy scale of small field inflation. We
recover the slightly disfavored values of sub-Planckian vacuum
expectation values. They clearly remain acceptable but only if the
reheating ends at a high energy. Again, the previous results are
compatible and easily understandable with the slow-roll predictions of
Sec.~\ref{subsec:smallfield}.

\par

To conclude this section, we have found that the current WMAP data
give non-trivial information on the reheating era in the two most
considered classes of prototypical inflationary models (large and small
fields). Conversely, when the goal is to use CMB data to constrain
these models, including the reheating into the marginalization is
definitely no longer an option.

\subsection{Importance sampling from slow-roll bounds}
\label{subsec:sampling}

The formulas derived in Sec.~\ref{subsec:why} link the reheating and
inflationary model parameters to the spectral index and
tensor-to-scalar ratio. As a result, they could be used to extract
constraints on the reheating energy scale from some already derived
constraints on the slow-roll parameters by using importance
sampling~\cite{Lewis:2002ah}. However, this method is highly
inefficient since most of the favored $(\nS,r)$ values do not
necessarily correspond to a consistent reheating model in a given
inflationary framework. However, it clearly illustrates how knowledge
on the primordial power spectra shape, complete with an inflaton
potential, translates into some information on the energy scale at
which the reheating ends. For this reason, we briefly discuss this
method in the following although we prefer an exact numerical
integration as performed above.

\par

Assuming an inflationary model, with a known potential $V(\phi)$, one
can generate any spectral index and tensor-to-scalar ratio such that,
at leading order in slow-roll,
\begin{equation}
\label{eq:srtoobs}
r = 16 \epsilon_{1*}, \qquad \nS - 1 = -2 \epsilon_{1*} - \epsilon_{2*},
\end{equation}
with
\begin{equation}
\label{eq:epstoobs}
  \epsilon_{1*} = \dfrac{1}{2} \left(\dfrac{V_*'}{V_*}\right)^2,
  \qquad \epsilon_{2*} = -2 \left[\dfrac{V_*''}{V_*} -
    \left(\dfrac{V_*'}{V_*} \right)^2 \right].
\end{equation}
A prime here is understood as a derivative with respect to the
field $\phi$. The star still refers to the time at which the scale
under consideration crossed out the Hubble radius during
inflation. Inverting Eq.~(\ref{eq:srtoobs}), or
Eq.~(\ref{eq:epstoobs}), gives the value of $\phi_*$ (and eventually
other parameters) leading to the required couple $(\nS,r)$. For
instance, in the large field models, one would find
\begin{equation}
\frac{\phi_*}{\Mp} = \sqrt{\dfrac{8 \epsilon_{1*}}{\epsilon_{2*}^2}}, 
\qquad p = 4 \dfrac{\epsilon_{1*}}{\epsilon_{2*}}\,.
\end{equation}
{}From $\phi_*$ one can derive $\Delta N_*$ from the slow-roll
trajectory and the value at which inflation stops, $\phi_\uend$. For
instance, for large field inflation, one would obtain
\begin{equation}
\Delta N_*=\frac{1-\epsilon_{1*}}{\epsilon_{2*}}.  
\end{equation}
The energy scale at which reheating ends stems from
Eqs.~(\ref{eq:Rrad}) and (\ref{eq:Rradw}):
\begin{equation}
\begin{aligned}
  \ln\left(\dfrac{\rhoreh^{1/4}}{\Mp} \right) &=\dfrac{3+3
    \bar{w}_\ureh}{1-3 \bar{w}_\ureh} (\Nzero + \Delta N_*) \\
  &- \dfrac{1+3 \bar{w}_\ureh}{2(1-3 \bar{w}_\ureh)} \ln\left(8 \pi^2
    P_*\right) + \ln \sqrt{\epsilon_{1*}} \\ &+
  \dfrac{1}{1-3 \bar{w}_\ureh} \ln\left(\dfrac{3}{\epsilon_{1*}}
    \dfrac{3-\epsilon_{1*}}{3 - \epsilon_{1\uend}}
    \dfrac{V_\uend}{V_*} \right),
\end{aligned}
\label{eq:srtorhoreh}
\end{equation}
provided $\bar{w}_\ureh\ne 1/3$. As expected, if the reheating is
radiation dominated, it cannot be distinguished from the usual
radiation era and $\rhoreh$ cannot be inferred. Clearly, for a given
set $(\epsilon_{1*}, \epsilon_{2*}, P_*)$, in an assumed model of
inflation, the right-hand side of Eq.~(\ref{eq:srtorhoreh}) is
uniquely determined and hence is $\rhoreh$. As already mentioned, such
a method is not well suited: picking up a random $(\epsilon_{1*},
\epsilon_{2*}, P_*)$ compatible with the power spectra shapes usually
predicts a value of $\rhoreh$ which is either incompatible with BBN, or
with the underlying model, \ie $\rhoreh > \rhoend$. The reason is
that inflationary physics is much more than Taylor expanding a
potential and fitting the power spectra shape. In order to solve this
issue, the way out is to perform an exact numerical integration of the
inflationary perturbations, including the reheating parameter, as we
previously did. This method has also the advantage to free ourselves
from any assumption on the equation of state parameter
$\bar{w}_\ureh$.

\section{Conclusion}
\label{sec:conclusion}

We now conclude our investigation by revisiting our main results. The
most important conclusion is that, both for large and small field
scenarios, the reheating parameter $\ln R$ is now constrained by CMB
data. The physical origin of this result is clear. For fixed physical
length scales today, a change in $\ln R$ modifies the location of the
CMB observable window along the inflationary potential, which is
possible only for a limited range of $\ln R$ given the data
accuracy. This conclusion is general and does not depend on the
details of the reheating epoch. However, if one assumes a model for
reheating, typically if one chooses a specific value of the mean
equation of state, then it becomes possible to express the constraints
mentioned above as limits on the energy density at the end of
reheating or, equivalently, as constraints on the reheating
temperatures. This leads to Eqs.~(\ref{eq:lfrhorehmin}).
and~(\ref{eq:conssf}). These results are of particular interest for
the supersymmetric extension of either large or small field
inflationary models. Indeed, our result limits the reheating
temperature from below whereas gravitinos production gives an upper
bound~\cite{Khlopov:1984pf, Kallosh:1999jj, Giudice:1999yt,
  Lemoine:1999sc, Maroto:1999ch, Giudice:1999am, Buonanno:2000cp,
  Copeland:2005qe,Jedamzik:2006xz, Kawasaki:2008qe, Bailly:2009pe}:
$T_\ureh < 10^{4}\,\TeV$, where $T_\ureh$ is given in
Eq.~(\ref{eq:Trehdef}). Therefore, assuming $g_*\simeq 200$ in the
large field case, we now have an allowed range of variation for the
reheating temperature given by
\begin{equation}
  6 \, \mbox{TeV}\lesssim T_\ureh\lesssim 10^4 \, \mbox{TeV}.
\end{equation}

\par

By including the reheating parameter in our analysis, we can
marginalize over all reheating history to infer the inflationary
parameter values in a robust way. For large field scenarios, we find
that the power index is upper limited by $p<2.2$, at $95\%$ confidence
limit. This means that the prototypical model of inflation, namely,
massive chaotic inflation, is now under pressure. Similarly, the small
field models with sub-Planckian vacuum expectation values $\mu$ are
slightly disfavored. In fact, $\mu$ is correlated with the potential
power $p$, as represented in Fig.~\ref{fig:sf2D}. Without
marginalization over $p$, large values of $p \gtrsim 6$ are actually
necessary to allow the sub-Planckian values of $\mu$ to be inside the
$95\%$ contour. If one has a theoretical prejudice for $\mu< \Mp$, and
for reasonable values of $p < 6$, then small field models can also be
considered under pressure.

\par

Since the constraints on the reheating parameters are directly related
to the ability of the data to determine the observable parameters, one
could consider more data sets than the WMAP7 data. In fact, since
solely the accuracy on the primordial parameters matters, data sets
improving the constraints on the standard cosmological parameters do
not change the reheating bounds. On the other hand, small scale CMB
experiments may be decisive but, as mentioned in
Ref.~\cite{Komatsu:2010fb}, they do not give a significant improvement
on the determination of $\nS$ and $r$ due to their low accuracy at
large multipoles. We have indeed tested that our limits do not change
by including the Baryonic Acoustic Oscillation~\cite{Percival:2009xn}
or the Arcminute Cosmology Bolometer Array Receiver
data~\cite{Reichardt:2008ay} in our analysis. On the other hand, since
the future Planck data are expected to improve the bounds on $\nS$,
$r$ and even on new primordial observables, we should get
unprecedented information on the inflationary reheating era.

\begin{acknowledgments}
  This work is partially supported by the Belgian Federal Office for
  Science, Technical and Cultural Affairs, under the Inter-university
  Attraction Pole Grant No. P6/11.
\end{acknowledgments}

\appendix

\section{Reheating from inflaton decay}
\label{appendix:lfreh}

For large field models, we have used the fact that
$\bar{w}_{\ureh}=(p-2)/(p+2)$, a well-known formula established for
the first time in Ref.~\cite{Turner:1983he}. However, this result
assumes that the inflaton field is not coupled to other fields, a
hypothesis that, if acceptable at the early stages of the reheating
phase, cannot be maintained if one is interested in the transition to
the radiation dominated era. The purpose of this Appendix is to take
into account the coupling of the inflaton field with the rest of the
world and to study its impact on the value of $\bar{w}_{\ureh}$. As we
show in the following, when the coupling is considered, the value of
$\bar{w}_\ureh$ does not significantly deviate from the equation given
above.
 
\par

\begin{figure}
\begin{center}
\includegraphics[width=0.5\textwidth,clip=true]{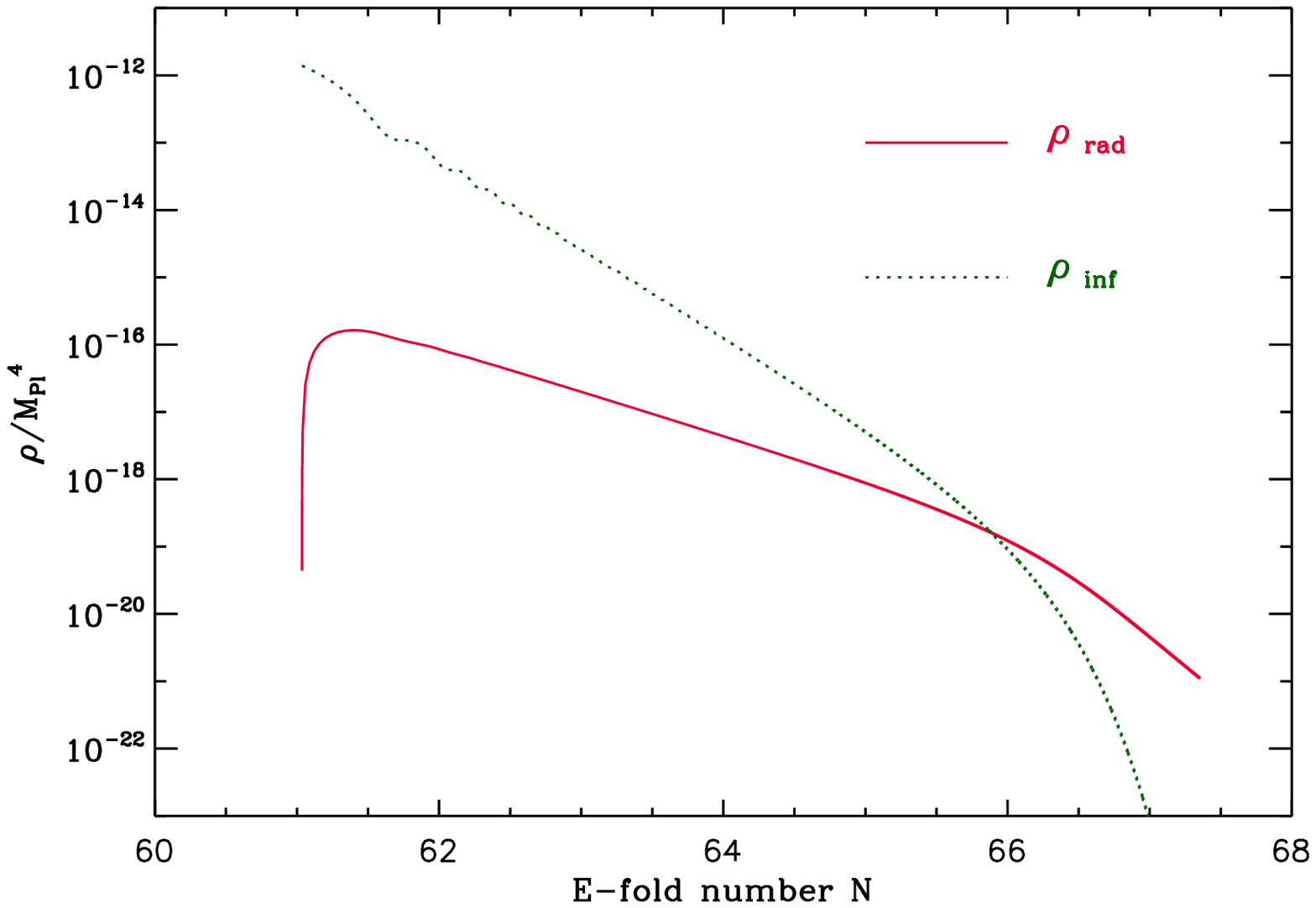}
\includegraphics[width=0.5\textwidth,clip=true]{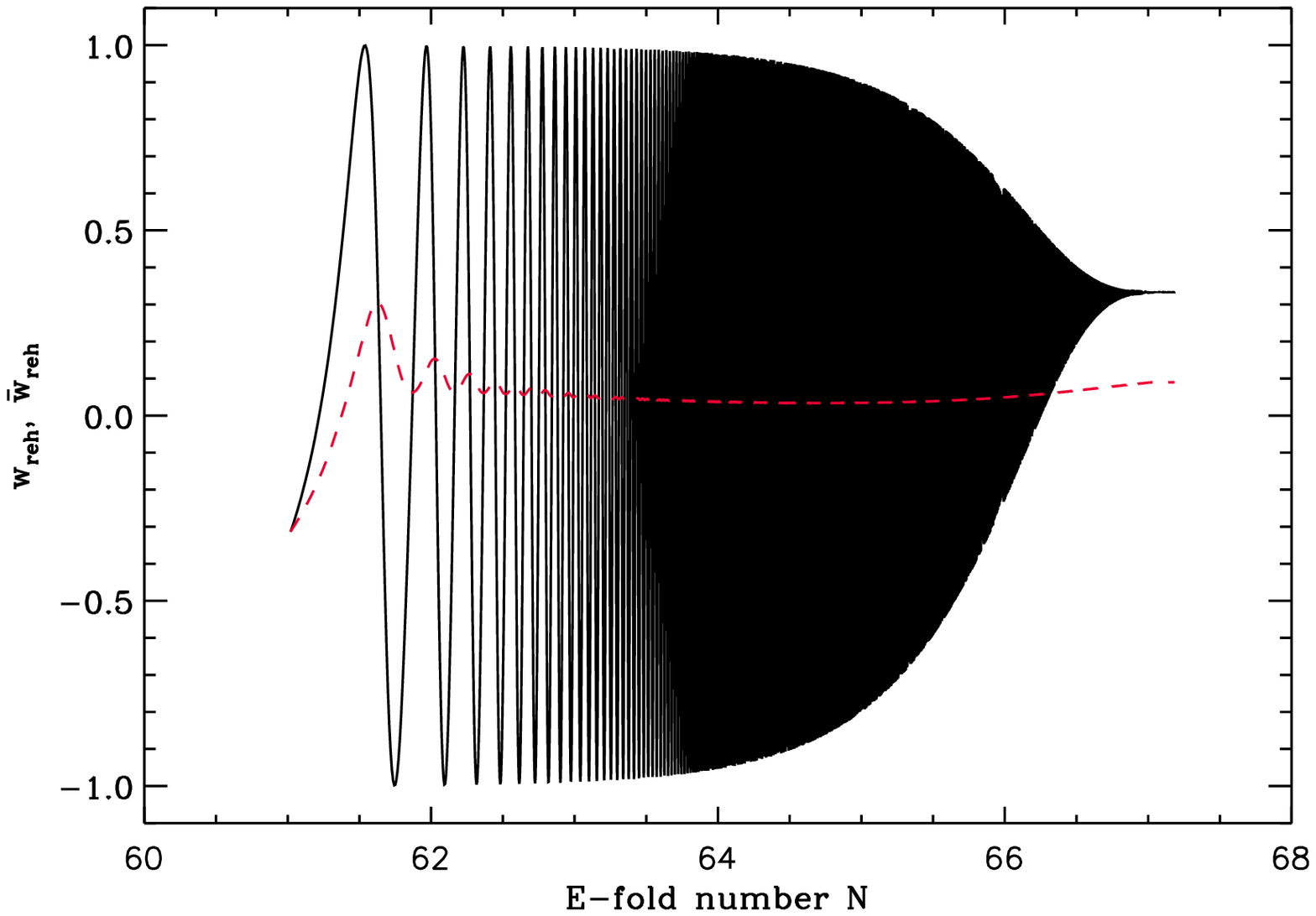}
\caption{Top panel: Evolution of the inflaton (dotted green line) and
  radiation (solid red line) energy densities during the reheating
  epoch as a function of the number of e-folds since the beginning of
  large field inflation with $p=2$ ($N_{_{\rm T}}\simeq61$). The
  inflaton decay rate has been chosen to be $\Gamma \simeq 1.375\times
  10^{9}$ $\GeV$ corresponding to a reheating temperature of
  $T_{\ureh}\simeq 3.2\times 10^{13}g_*^{-1/4}\,\GeV$. The total
  number of e-folds during the reheating epoch is $N_\ureh=5$. Bottom
  panel: the instantaneous equation of state (black solid line),
  $w_{\ureh}=(P_{\rm inf}+\rho_{\gamma}/3)/(\rho_{\rm
    inf}+\rho_{\gamma})$, during the reheating era and going to
  $1/3$. The dashed red line represents $\bar{w}_\ureh$. Its value
  when the instantaneous equation of state has reached $1/3$ is
  $\simeq 0.08$ so that, even when the number of e-folds during
  reheating is small, deviations from $\bar{w}_\ureh=0$ never exceed
  $8\%$.}
\label{fig:reheatmodel}
\end{center}
\end{figure}

A simple and standard way to model the physical situation where the
inflaton field decays into radiation is to write the Klein-Gordon
equation as~\cite{Turner:1983he}
\begin{equation}
\label{eq:KGmodified}
\ddot{\phi}+(3H+\Gamma)\dot{\phi}+\frac{{\rm d}V}{{\rm d}\phi}=0,
\end{equation}
with $\Gamma $ being the inflaton decay rate. The evolution equation for
the radiation energy density is modified accordingly in order to
ensure that the total energy density is conserved. One obtains
\begin{equation}
\label{eq:radconsmodified}
\frac{{\rm d}\rho _{\gamma }}{{\rm d}N}+4\rho_{\gamma }
=\frac{2p}{p+2}\frac{\Gamma }{H}\rho_{\rm inf},
\end{equation}
where $\rho_{\rm inf}\equiv \dot{\phi}^2/2+V(\phi)$. Of course, these
two formulas must be supplemented with the Friedmann--Lema\^{\i}tre
equation, $H^2=\left(\rho_{\rm inf}+\rho_{\gamma}\right)/(3\Mp^2)$. We
have numerically integrated Eqs.~(\ref{eq:KGmodified})
and~(\ref{eq:radconsmodified}) for large field inflation with $p=2$
and the corresponding evolutions of $\rho _{\rm inf}$, $\rho_{\rm
  rad}$, $w_\ureh$ and $\bar{w}_\ureh$ are displayed in
Fig.~\ref{fig:reheatmodel}.

\par

There are some subtle issues if one wants to numerically evaluate the
value of $\bar{w}_{\ureh}$. For instance, the time at which one
considers the Universe to be reheated is not very well defined since
we have a smooth transition, and this affects the precise numerical
determination of $\bar{w}_{\ureh}$. Indeed, one could consider that
reheating is completed when $t\sim \Gamma ^{-1}$ or when
$w_\ureh=1/3\pm \delta$ for a given $\delta $. These choices (for
instance the precise value of $\delta$) lead to different values of
the mean equation of state. Moreover, the numerical calculation itself
can be difficult since one has to integrate a rapidly oscillating
function. All in all, we find that, for $N_{\rm reh}\simeq 5$ (as in
Fig.~\ref{fig:reheatmodel}), $\bar{w}_{\ureh}\lesssim 0.08$, that is
to say a value close to $0$. Let us also notice that when the value of
$N_\ureh$ increases (\ie when the reheating temperature decreases),
one expects this value to be even less than the limit quoted before
(but this regime is numerically difficult to follow since the equation
of state rapidly oscillates during a long time). Therefore, we
conclude that Eq.~(\ref{eq:wrehlf}) can reasonably be trusted and this
justifies the approach used in this paper.

\begin{figure}
\begin{center}
\includegraphics[width=8.6cm]{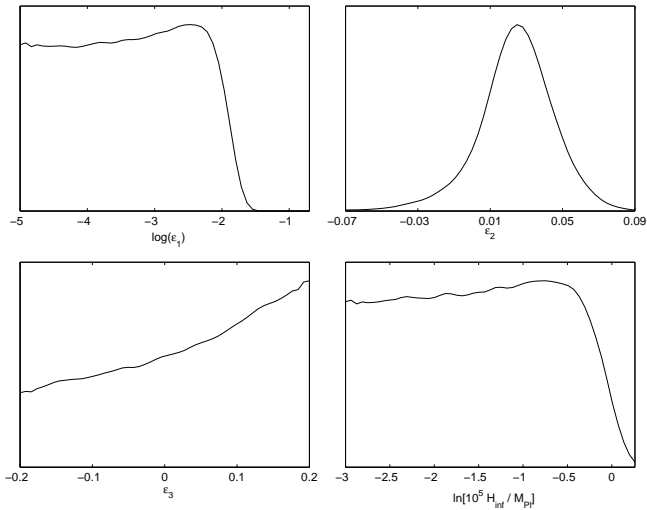}
\caption{Marginalized posterior probability distributions (solid
  lines) and mean likelihoods (dotted lines) for the three first slow-roll
  parameters and for the energy scale of inflation ($\kappa \equiv
  1/\Mp$).}
\label{fig:sfpostsr}
\end{center}
\end{figure}
\begin{figure}
\begin{center}
\includegraphics[width=8.8cm]{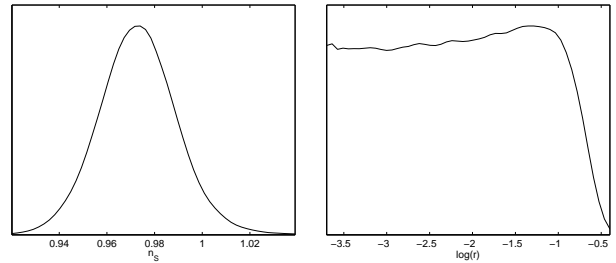}
\caption{Marginalized posterior probability distributions (solid
  lines) for the spectral index and the tensor-to-scalar ratio in the
  slow-roll approximation. The primordial power spectra are expanded
  at second order in slow-roll.}
\label{fig:sfpostnsR}
\end{center}
\end{figure}

\section{Slow-Roll posterior distributions}
\label{appendix:srpost}

In Sec.~\ref{sec:physicalorigin}, we used the one- and two-sigma
contours in the planes $(\nS,r)$ and $(\epsilon_1,\epsilon_2)$ obtained
from the WMAP7 data. In this section, for the sake of completeness, we
also provide the resulting one-dimensional marginalized posterior
probability distributions on those four parameters together with the
third slow-roll parameters $\epsilon_3$ and the energy scale of
inflation. Let us notice that our analysis, even for $\nS$ and $r$,
assumes the second order slow-roll expanded primordial power spectra
for both the scalar and tensor perturbations.

In Fig.~\ref{fig:sfpostsr}, we present the marginalized posterior
probability distributions for the three first slow-roll parameters and
for the Hubble parameter during inflation ($H_\uinf = H_*$). For the
first slow-roll parameter, we have assumed a Jeffreys' prior in the
range $[10^{-5},0.2]$, \ie a flat prior on $\log(\epsilon_1)$ in
the range $[-5,-0.7]$, as appropriate for a parameter the order of
magnitude of which is unknown. For the two next slow-roll parameters,
we have chosen flat priors in $[-0.2,0.2]$. The upper limit on
$\epsilon_1$ directly comes from the level of primordial
gravitational waves. The parameter $\epsilon_2$ is well constrained
while $\epsilon_3$ remains unbounded. There exists an upper limit on
the energy scale of inflation which, as for the first slow-roll
parameter, directly comes from the non-observation of the primordial
gravitational waves. In Fig.~\ref{fig:sfpostnsR}, the spectral index
and the tensor-to-scalar ratio have been plotted and obtained by
importance sampling from the slow-roll constraints. Indeed, at second
order in slow-roll, one has
\begin{equation}
\begin{aligned}
  \nS &= 1-2\epsilon_1-\epsilon_2-2\epsilon_1^2-(2C+3)\epsilon_1\epsilon_2
-C\epsilon_2\epsilon_3 \\
r &= 16\epsilon_1\Biggl[1+C\epsilon_2+\left(C-\frac{\pi^2}{2}+5\right)
\epsilon_1\epsilon_2  \\ 
&+ \left(\frac{C^2}{2}-\frac{\pi^2}{8}+1\right)\epsilon_2^2
+\left(\frac{C^2}{2}-\frac{\pi^2}{24}\right)\epsilon_2\epsilon_3\Biggr],
\end{aligned}
\end{equation}
where $C$ is a numerical constant, $C\simeq -0.7296$. As can be seen
on these plots, the spectral index value $\nS=1$ is disfavored but
not excluded when marginalizing over second order slow-roll. As for
$\epsilon_1$, there is only an upper limit on the tensor-to-scalar
ratio $r$.

\bibliography{biblio}
\end{document}